\UseRawInputEncoding

\documentclass[prd,aps,showpacs,nofootinbib]{revtex4}
\usepackage{amssymb}
\usepackage{graphicx}
\usepackage{amsmath}

\newcommand{\be}{\begin{equation}}
\newcommand{\ee}{\end{equation}}
\newcommand{\bq}{\begin{eqnarray}}
\newcommand{\eq}{\end{eqnarray}}

\begin{document}
\title{Cosmology of the Symmetrical Relativity versus Spontaneous Creation of the Universe Ex Nihilo} 

\author{*Cl\'audio Nassif Cruz and Fernando Ant\^onio da Silva}

\affiliation{{\bf *CPFT-MG}: Centro de Pesquisas em F\'isica Te\'orica-MG, Rua Rio de Janeiro 1186, Lourdes, Belo Horizonte-MG, CEP:30.160-041, Brazil.\\
*claudionassif@yahoo.com.br; fernandosilvaantonio16@gmail.com} 

\begin{abstract}
The cosmology of ``Spontaneous Creation of the Universe Ex Nihilo''\cite{CEN} and the cosmology of 
the Symmetrical Relativity\cite{TVSL} offer proposals to explain the creation and evolution of the universe. In essence they are still very distinct. However, we will argue that there was an antecedent to the big bang. Thus, we will penetrate a trans-Planckian regime, where we find an effective Planck length $L_P^{\prime}\rightarrow 0$. This will lead to more fundamental physical reflections about {\it nothing} in the cosmology of spontaneous creation\cite{CEN}. From the point of view of spontaneous creation, the step backwards went back to another principle, based on the {\it information}, which led to the big bang. While the spontaneous creation refers to the virtual pre-existence of {\it information}, which would have emerged randomly from {\it nothing}\cite{CEN}, the cosmology of Symmetrical Relativity\cite{TVSL} does not stop there: we go back to one's own origin by projecting it before the creation of one's own time. From the present perspective, {\it nothing} is a primordial vacuum, whose {\it information} has made the universe by condensing and igniting. It was not randomly created, since the entropy had to vary from infinite (chaos) to zero (big bang) by violating the 2nd.law of thermodynamics in a trans-Planckian regime. {\it Nothing} or chaos with infinite entropy precedes the big bang (null entropy) within a trans-Planckian scenario and it determines the whole plot until the total extinction of the universe. We will show that {\it information}, besides not being born of the universe, also does not develop from it like the computational idea of Artificial Intelligence (AI). Thus, the universe is not simply self-taught as defended by the spontaneous creation. This could shed light on the problem of Penrose's Weyl curvature hypothesis, which considers a null Weyl curvature due to a null entropy in the big bang. 
\end{abstract}

\maketitle

\section{Introduction} 

Driven by the search for new fundamental symmetries in nature, the so-called Symmetrical Special Relativity (SSR)\cite{N2016} attempts to implement a uniform background field in the whole spacetime, and it is the basis of Cosmology of the Symmetrical Relativity\cite{TVSL}. Such a background field connected to a uniform vacuum energy density represents the preferred reference frame $S_V$\cite{N2016}, which led to postulate the universal minimum limit of speed $V$ for particles with very large wavelengths (very low energies).

The idea that some symmetries of a fundamental theory of quantum gravity may have non trivial consequences for cosmology and particle physics at very low energies is interesting and indeed quite reasonable. So it seems that the idea of a universal minimum speed as one of the first attempts of Lorentz symmetry violation could have the origin from a fundamental theory of quantum gravity at very low energies (very large wavelengths). Besides quantum gravity for the Planck minimum length $L_P$ (very high energies), the new symmetry idea of a minimum speed $V$ could appear due to the indispensable presence of gravity at quantum level for particles with very large wavelengths (very low energies). So we expect that such a universal
minimum speed $V$ also depends on fundamental constants as for instance the constant of gravitation $G$ and the Planck constant $\hbar$, as already shown in a previous paper\cite{N2016}. In this sense, there could
be a relationship between $V$ and $L_P$, since $V\propto L_P$\cite{N2016}. 

The hypothesis of the lowest non-null limit of speed $V$ for very low energies ($V\approx v<<c$) in the 
spacetime results from the following physical motivation:

- In non-relativistic quantum mechanics, the plane wave wave-function ($Ae^{\pm ipx/\hbar}$) which represents a free particle is an idealisation that is impossible to conceive under physical reality. In the event of such an idealized plane wave, it would be possible to find with certainty the reference frame that cancels its momentum ($p=0$), so that the uncertainty on its position would be $\Delta x=\infty$. However, the presence of an unattainable minimum limit of speed $V$ emerges in order to prevent the ideal case of a plane wave wave-function ($p=constant$ or $\Delta p=0$). This means that there is no perfect inertial motion ($v=constant$) such as a plane wave, except the privileged reference frame of a universal background field connected to such unattainable minimum limit of speed $V$, where $p$ would vanish. However, since such a minimum speed $V$ (universal background frame $S_V$) is unattainable for the particles with low energies (large length scales), their momentum can actually never vanish when one tries to be closer to such a preferred frame ($S_V$) that represents a fundamental zero-point energy for $v\rightarrow V$. 

 On the other hand, according to Special Relativity (SR), the momentum cannot be infinite since the maximum speed $c$ is also unattainable for a massive particle, except the photon ($v=c$) as it is a massless particle.

 This reasoning allows us to think that the electromagnetic radiation (photon: $``c-c''=c$) as well as the massive particle ($``v-v''>V$ for $V<v<c$)\cite{N2016} are in equal-footing in the sense that it is not possible to find a reference frame at rest for both through any speed transformation in the spacetime of SSR.

 The conception of {\it information} must be indeed implicated in the logical chain of natural events or processes. Otherwise, there would be no possible knowledge. There is no way to do without it. However, what is at issue is not the information itself, but its origin. By means of the variation of the speed of light and a minimum speed in the scenario of an inflationary universe with accelerated expansion (cosmology of the Symmetrical Relativity)\cite{TVSL}, we have realized that a process was under way even before the big bang (a trans-Planckian regime). The initial conditions were not originated, but only situated at this special moment of creation, including the creation of one's own time.

We emphasize the importance of time in the creation of the universe, because of what we have to reveal later.

We agree with the hypothesis that there was a precedent movement into big bang, but not devoid of {\it cause}, as the authors of the spontaneous creation\cite{CEN} authors would wish. What seems to us to be {\it chance} is the power of an implicit determinism. Therefore neither {\it chance} nor merely {\it information} within the scenario of a kind of AI was responsible by the big bang. 

Spontaneous creation ex nihilo\cite{CEN} brought an interesting embryonic conception. However, it is unsustainable given its lack of mathematical rigor. No wonder it makes an impression. Just the idea of ​​basing it on {\it information} - the idea is not even original- is good, but as long as it is merely {\it information}, because it is necessary in the context of random choices.

There is a kind of tacit agreement about the belief in the existence of {\it nothing} or a primordial vacuum of pre-big bang (a trans-Planckian vacuum shown in Fig.5,a). Such an idea is common to the cosmology of the spontaneous creation\cite{CEN} and the cosmology of the symmetrical relativity\cite{TVSL}, but we note differences in their proposals regarding the views on formalism and rigor that are absent in the concurrent research of the spontaneous creation. This prevents spontaneous creation and puts cosmology of the symmetrical relativity from side by side in the face of obvious antagonism between them with respect to how the primordial vacuum (trans-Planckian vacuum) led to the emergence of the big bang and the dark energy as a Planckian (cosmic) vacuum, which is responsible by the cosmic inflation and the accelerated expansion of the universe. 

The famous question related to Penrose's Weyl curvature hypothesis in the big bang will be better clarified 
by the cosmology of the symmetrical relativity in the last section. 

\section{A brief review of SSR}

Symmetrical Special Relativity (SSR)\cite{N2016}\cite{N2015}\cite{N2008}\cite{Rodrigo}\cite{N2018}\cite{N2010}\cite{N2012}\cite{tachyon}\cite{uncertainty} is essentially an extension of Special Relativity (SR) by correcting SR in the regime of very low energies just for subatomic particles (quantum world), so that one postulates the existence of an invariant minimum speed $V$ with the same status of the invariance of the speed of light $c$ for very high energies of a massive particle, i.e., we have $V<v<c$ for any massive particles. 

In SSR-theory, the invariant minimum speed $V$ is associated with a zero-point energy connected 
to a fundamental vacuum energy of gravitational origin\cite{N2016}. As such vacuum energy has
a kinematic origin (an invariant minimum speed), this changes the causal structure of spacetime,
so that we are led to alter our conception of reference frames by introducing a preferred 
(absolute) reference frame in the spacetime. This preferred reference frame is associated with the invariant minimum speed $V$ and thus being a unattainable frame for any massive particles when one tries to reduce their energies to zero close to such absolute frame. Therefore, it becomes impossible to place exactly any particles at such preferred reference frame so-called {\it ultra-referential} $S_V$\cite{N2016}.

 SSR-theory was able to predict the origin of the tiny positive value of the cosmological 
 constant $\Lambda$ connected to the very low energy density of
 vacuum\cite{N2016}\cite{N2015}\cite{N2008}. It was also able of obtaining of the equation of state (EOS) of vacuum $p(pressure)=-\rho(vacuum~energy~density)$\cite{N2016}\cite{N2015}\cite{N2008}, which represents the cosmological constant being related to the preferred reference frame or the ultra-referential $S_V$ of SSR, thus leading to the cosmological anti-gravity or the accelerating expansion of the universe. 
 
 SSR-theory has many other successfully predictions as, for instance, a fundamental explanation 
 for the uncertainty principle\cite{N2012} within a metric 
 scenario\cite{uncertainty} of type of de-Sitter (dS) space, since we have already shown that 
 SSR-metric is a conformal metric\cite{Rodrigo}. We can still mention other good predictions 
 of the theory as the question of the third law of thermodynamics, which leads to a minimum temperature $T_{min}$\cite{N2018} in the cosmological scenario of the distant future of the
 universe\cite{TVSL}.

\begin{figure}
\includegraphics[scale=0.9]{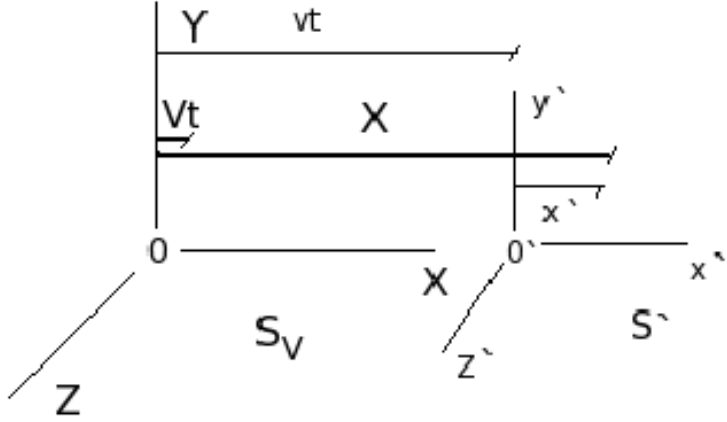}
\caption{In this special case $(1+1)D$, the referential $S^{\prime}$ moves in $x$-direction with a speed $v(>V)$ with respect to a preferred reference frame connected to the ultra-referential $S_V$ (background field). If $V\rightarrow 0$, $S_V$ is eliminated and thus the Galilean frame $S$ takes place, recovering the Lorentz transformations.}
\end{figure}

 The $(1+1)D$-transformations in SSR with $\vec v=v_x=v$ (Fig.1) are 

\begin{equation}
x^{\prime}=\Psi(X-vt+Vt)=\theta\gamma(X-vt+Vt) 
\end{equation}

and 

\begin{equation}
t^{\prime}=\Psi\left(t-\frac{vX}{c^2}+\frac{VX}{c^2}\right)=\theta\gamma\left(t-\frac{vX}{c^2}+\frac{VX}{c^2}\right), 
\end{equation}
where the factor $\theta=\sqrt{1-V^2/v^2}$ and $\Psi=\theta\gamma=\sqrt{1-V^2/v^2}/\sqrt{1-v^2/c^2}$. We have found that $V=\sqrt{Gm_pm_e/4\pi\epsilon_0}(q_e/\hbar)
\approx 4.58\times 10^{-14}$m/s\cite{N2016}, where $m_p$ is the proton mass, $m_e$ is the electron mass and $q_e$ is the electron charge. 

\begin{figure}
\includegraphics[scale=1.0]{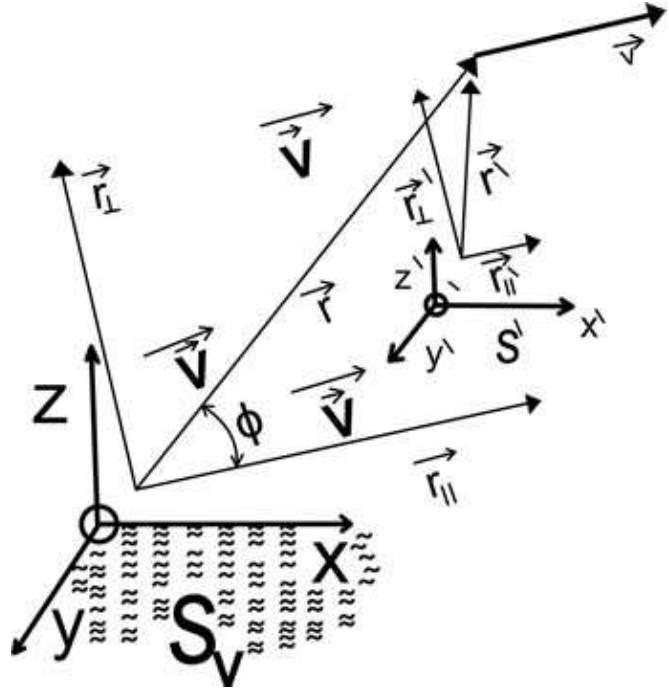}
\caption{$S^{\prime}$ moves with a $3D$-velocity $\vec v=(v_x,v_y,v_z)$ in relation to $S_V$. For the special case of $1D$-velocity
$\vec v=(v_x)$, we recover Fig.2; however, in this general case of $3D$-velocity $\vec v$, there must be a background vector $\vec V$ (minimum velocity) with the same direction of $\vec v$ as shown in this figure. Such a background vector $\vec V=(V/v)\vec v$ is related to the background reference frame (ultra-referential) $S_V$, thus leading to Lorentz violation. The modulus of $\vec V$ is invariant at any direction.} 
\end{figure}

The $(3+1)D$-transformations in SSR (Fig.2) were shown in a previous paper\cite{N2016}, namely: 

\begin{equation}
\vec r^{\prime}=\theta\left[\vec r + (\gamma-1)\frac{(\vec r.\vec v)}{v^2}\vec v-\gamma\vec v(1-\alpha)t\right], 
\end{equation}
where $\alpha=V/v$. 

And 

\begin{equation}
t^{\prime}=\theta\gamma\left[t-\frac{\vec r.\vec v}{c^2}(1-\alpha)\right]. 
\end{equation}

If we make $V\rightarrow 0$, we recover the well-known Lorentz transformations.

From Eq.(3) and Eq.(4), we can verify that, if we consider $\vec v$ to be in the same direction of $\vec r$, with $r=X$, we recover the special case of $(1+1)D$-transformations given by Eq.(1) and Eq.(2). 

The metric of SSR is a Minkowski metric deformed by a multiplicative function, i.e., a scale factor $\Theta(v)$ with $v$-dependence. $\Theta(v)$ works like a conformal factor, which leads to a kind of dS-metric\cite{Rodrigo}, namely $d\mathcal S^{2}=\Theta\eta_{\mu\nu}dx^{\mu}dx^{\nu}$, where $\Theta=\Theta(v)=\theta^{-2}=1/(1-V^2/v^2)\equiv 1/(1-\Lambda r^2/6c^2)^2$\cite{Rodrigo} is the conformal factor and
$\eta_{\mu\nu}$ is the well-known Minkowski metric. 

\section{Deformed Symmetrical Special Relativity: Origin of the Variation of the Speed of Light or Cosmology of the Symmetrical Relativity} 

The variation of the speed of light $c$ and the minimum speed $V$ with the temperature of the expansion universe leads to a Deformed Symmetrical Special Relativity (DSSR)\cite{TVSL}, by setting its limits as boundary conditions, for the beginning and the end of time. These speed limits relate respectively to $T_P$ (maximum temperature of the big bang so-called Planck temperature) and $T_{min}$ (minimum temperature of the big rip), which corresponds to the initial inflationary period known as Alan Guth period and the final inflationary period to be verified at the moment of the death of the Universe, according to DSSR, being estimated for some billions of years from now on.

By the way, the cosmology of the Symmetrical Relativity\cite{TVSL} is the only one that shows that the extinction of the universe will not happen by thermal death, as predicted by all the rest of the mainstream cosmologies.

Around 1850, William Thomson (Lord Kelvin) extrapolated the view of the loss of mechanical energy in the form of heat - according to the first law of thermodynamics - to the death condition of the universe. His point of view has been readily adopted {\it en passant} by cosmologists ever since. By following the guidance imposed by high energy physics, according to the Theory of Relativity, all the physicists turned their gaze to the initial big bang inflationary moment and left cosmology of the Symmetrical Relativity\cite{TVSL} to discover the big rip as the final inflation of the universe.

The limits of temperature $T_P$ and $T_{min}$ define the so-called Planck temperature range\cite{N2018}. It represents the spectrum of physically possible temperatures. The Kelvin absolute temperature scale - from zero to infinite - in no way corresponds to the cosmological scenario\cite{N2018}. But if we think about it, it was even convenient to denominate it as being absolute, as its limits of temperature ($0<T<\infty$) are ideal (not physical) or without any connection with the cosmological scenario of temperatures shown in ref.\cite{N2018}. 

\begin{figure}
\includegraphics[scale=1.0]{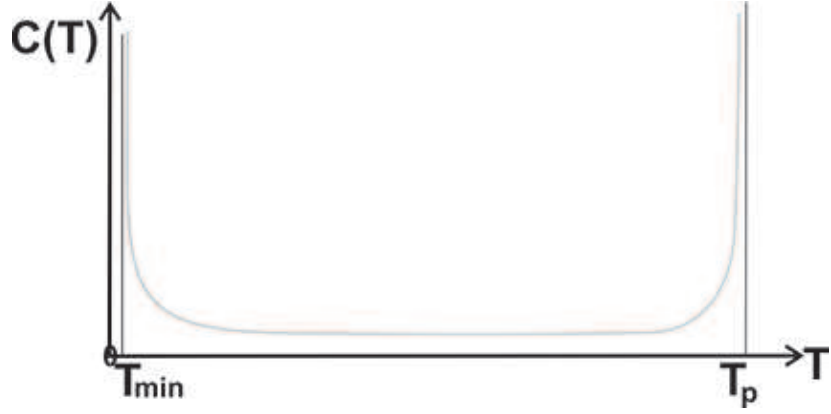}
\caption{This graph for representing Eq.(5) shows that the speed of light diverges for both
limits of temperature, namely Planck temperature $T_P(\sim 10^{32}K)$ for the scale of 
Planck $L_P(\sim 10^{-35}m)$ in the early (too hot) universe, and a minimum temperature 
$T_{min}(\sim 10^{-12}K)$ in a ultra-cold universe connected to a horizon radius $r_h>>r_u(\sim 10^{26}m)$.} 
\end{figure}

The constants $V$, $c$, the Planck temperature $\mathcal T_P$ and $T_{min}$ appear in the function
$c(T)$\cite{TVSL}, as follows: 

\begin{equation}
c(T)=c^{\prime}=\frac{c}{\sqrt{1-\frac{T}{\mathcal T_P}}\sqrt{1-\frac{T_{min}}{T}}}. 
\end{equation}

This function is extremely important as it allows us to draw the full picture of the evolution of the universe, from the big bang to the big rip. On the other hand, the reciprocal function of Eq.(5) provides the minimum speed with temperature dependence, namely: 
$V(T)=V^{\prime}=V\sqrt{1-T/\mathcal T_P}\sqrt{1-T_{min}/T}$\cite{TVSL}. 

\begin{figure}
\includegraphics[scale=0.60]{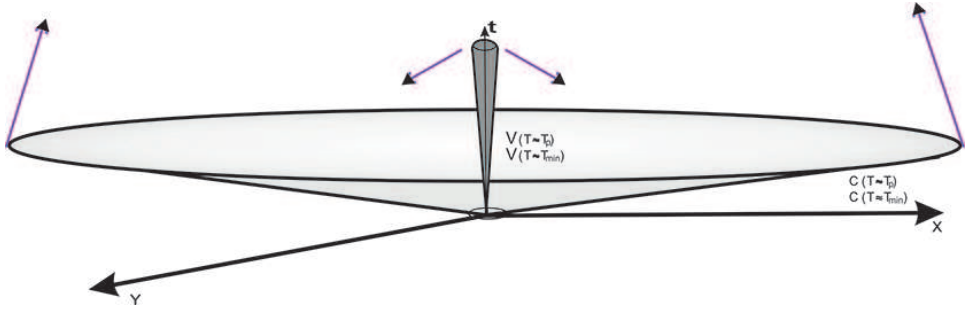}
\caption{The figure represents the dark cone and light cone. They are opposite aspects of a same
transcent state that works like a Newtonian space given in Fig.6, i.e., a flat space related to a primordial vacuum, where $c^{\prime}\rightarrow\infty$ and $V^{\prime}\rightarrow 0 $ for both limits of temperatures, namely $T_P$ and $T_{min}$, as shown in Eq.(5).} 
\end{figure}

\section{Spontaneous Creation Ex Nihilo}

As we have been repeating, the spontaneous creation ex nihilo\cite{CEN} and cosmology of the Symmetrical Relativity\cite{TVSL} are not limited to the cosmic phase, or the phase that began with the big bang. However, the authors of the spontaneous creation intend to probe the inscrutable by making exclusive use of a dialectical method by using the computational language of information, i.e., bit $(+)$ and bit $(-)$. 

The authors of the spontaneous creation\cite{CEN} acknowledge that they have formulated a vague proposal. They themselves admitted the informality of their paper and declared the lack of the necessary rigor. So, was it not premature to introduce it?

However, they are right at a first view when they claim that the spontaneous creation adjusts to state-of-the-art cosmologies starting from the big bang. But, with the cosmology of the Symmetrical Relativity\cite{TVSL}, we claim that the spontaneous creation ex nihilo\cite{CEN} is already one step ahead of the big bang. However, the authors of the spontaneous creation have conjectured that {\it chance} had been the cause of the unique symmetry breaking species of so-called NIEs within {\it nothing}. {\it Nothing} would be the uncreated spacetime (the primordial vacuum so-called trans-Planckian vacuum), which can be activated virtually by the {\it information} that led firstly to a primordial {\it big crunch} in order to generate the initial singularity within the Planck scale with $T_P$, thus leading to the big bang. 

The supposed and unintelligible {\it symmetry breaking} of the so-called {\it information} (NIEs)\cite{CEN} induced, in addition to materializing in the {\it big crunch}, had caused entropy to rise almost to infinite and provided the initial conditions for the big bang, according to the authors of the spontaneous creation, which opposes the cosmology of the Symmetrical Relativity. Their purpose was not to circumvent the intriguing fact that the entropy had to decrease from infinite (the absolute chaos of a primordial vacuum) to zero, by creating order (zero entropy in the big bang with the dark energy) from such chaos of a primordial vacuum (trans-Planckian vacuum), whereas the cosmology of the Symmetrical Relativity is able to do that, as 
it will be shown. 

\section{Comparasion Points}

From the perspective of SSR and cosmology of the Symmetrical Relativity, instead of {\it chance} in the spontaneous creation, we will perceive an implicit determinism giving rise to the creation (big bang). 
 
Certainly, the enthusiasm of the authors of spontaneous creation was heightened, as they saw oppositional relationships, which are presented in various theoretical models. For it seems that they are unaware that such relationships underlie any reasoning because they are inherent in physical systems and processes in general. Thus, they are obviously necessary.

We are referring to the examples selected by them for the purpose of illustrating their article\cite{CEN}: the problems of parity (matter and anti-matter), pair production, baryonic and anti-baryonic matter, etc.

They\cite{CEN} mentioned them for the purpose of suggesting resonances to their ideas.

Presenting the universe as a self-ignited machine by {\it information} - subject to such a kind of {\it self-programming}, which determines its own evolution - bears analogies with the dichotomy between the computer (hardware) and its operating system (software), beyond Artificial Intelligence (AI): the universe would have formed spontaneously out of {\it nothing}\cite{CEN} and, by mistake and trial, {\it change} would lead to learn with itself the steps of evolution. The universe had to pass endless tests of {\it russian roulette}. It is unbelievable that it survived the last $15$ billion years, if in the fleeting instant of each quantum transition it could simply explode. Imagine that for such a long time and during every very brief moment of a quantum transition, in every attempt with the dice always obtain the face of number $6$.

According to the establishment, a transition is in many respects equivalent to EPR quantum transmissions: the electron transports from one layer of the electrosphere to another. If so, the facts of the electron disappearing from a point $A$, from the layer it was in, and appearing at a point $B$, situated on the other layer where it will be, are simultaneous events. Simultaneous events would in fact imply the acausality between them, since neither can be a cause of each other, since it is even necessary that a fact occurring at a time $t_{n+1}$ be preceded by another at time $t_n$, in order to ensure that there is a causal relationship between them. We understand Einstein's despair over these impertinent statements.

Indeed, the central idea of ​​the spontaneous creation- the {\it infoelements} or bits that process {\it information} emerge from {\it nothing} and then organize everything from the big bang to the end- no doubt comes from analogies with science of information to some extent acceptable. 

In this scenario, in which we perceive a kind of {\it cosmic program} driving the cosmology of the Symmetrical Relativity, both the machine (the universe) and its {\it operating system} (software in the context of information itself) are still due to an {\it uncertain cause}, far from being clarified by the 
spontaneous creation\cite{CEN} because arguments and evidence have not been pointed out, nor there is so reasonableness in that paper. The spontaneous creation proposal is abstract in the sense of being more qualitative or without a rigorous mathematical formalism as given by the cosmology of the Symmetrical Relativity and its thermodynamics implications in the cosmological scenario of the early universe. 

Information underlies natural processes and without it there would be no form of organization that could overcome chaos. There is no arguing. But to suppose that {\it information} proceeds from {\it nothing} is excessively improbable. 

Cosmology of the Symmetrical Relativity\cite{TVSL} reveals that light was of utmost importance in creation and its variation of speed determines the evolution and will determine the extinction and death of the universe, which will occur by the intriguing action of temperatures as high as those of the big bang.

We already know that there is a tremendous amount of {\it information} contained in the internal arrangement of atoms in the cross-linked solids (which determines crystal growth), as well as {\it information} that leads to stelar nucleosynthesis; innate information that also determines the special properties of the electron and other particles. 

This gives rise to some intriguing unsolved questions, namely: 

1) Where does the {\it information} come from?

2) How were they programmed?

Note that spontaneous creation's alleged solution brings further doubt and does not eliminate the original. By predisposing to adopt it, the subject will only have more questions about what to think and less to conclude. 

Recognizing that {\it chance} is the primary cause like spontaneous creation, or that an implicit determinism has provided input to the entire process of conception, creation, evolution and destiny of the universe such as the cosmology of the Symmetrical Relativity, thus how can we decide this question of origins, which is of the utmost importance, of which there is nothing concrete to offer any certainty? We are bound to settle for hypothesis. So, how do you get a minimum of conviction without analyzing each perspective as seriously and deeply as possible? In the end, each person will decide for him (her) self, convinced or not, for the arguments that may arise through theories, which can only be based on indirect ideas and observations, never on a fact in themselves, because they can never be directly tested. For every time you point out as first cause something as vague as {\it information}, the question will not still silence: but what caused it? 

 As the software/hardware distinction, the authors of the spontaneous creation propose the {\it information/universe} dichotomy. In its conception, {\it information} is pre-existing to cosmic processes and competes for them in the same direction, but on a different track, until they intersect at point $A$ and/or point $B$ (in the case of cosmologies that accept or dispense the inflation). Both from point $A$ and point $B$ (Fig.4), the two trajectories follow the same path, thus indicating that the universe began to manage its own {\it information}, being in analogy with Artificial Intelligence (AI). It would be a kind of machine produced from {\it nothing} due to {\it information}, which also appeared previously from nowhere, then merging 
 {\it information} and universe in a self-sustaining way.

We see that the problem of the origin of the universe was not even addressed at CEN, but only transferred to {\it information}. It became the cause of input in the cosmology of the spontaneous creation ex nihilo.

i) But what provides the input to {\it information} ex nihilo?

ii) What is {\it nothing} or the primordial vacuum in the scenario of the spontaneous creation?

We should substantially advance this issue by considering the cosmology of the Symmetrical Relativity.

\begin{figure}
\includegraphics[scale=0.3]{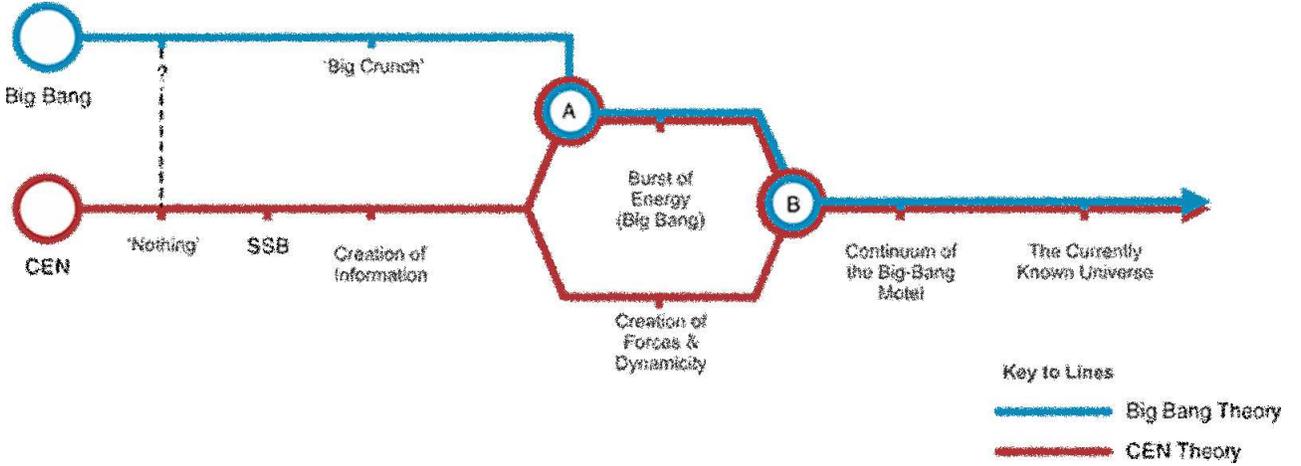}
\caption{A schematic route map of the big bang theory and the theory of the spontaneous creation. The blue time-line represents major milestones addressed by the big bang theory, and the red line represents those of the spontaneous creation theory. Joints $A$ and $B$ point out alternative merge points.}
\end{figure}

\section{The scalar fields (potentials) of SSR}

SSR-theory reveals that the hadronic matter and energy are responsible for the gravity field and the dark energy is responsible for the anti-gravity field. The intensity of both fields also depends on the temperature of the Background Cosmic Radiation (BCR), always in perfectly dialectical relationships. When the temperature corresponds to the high energy region, gravity predominates because of its direct dependence on it; and when the temperature corresponds to very low energy region, anti-gravity predominates also because of its direct dependence on it. High temperatures activate gravity when the hadrons are overheated while disabling anti-gravity. On the other hand, low temperatures activate anti-gravity when the ultra-cold dark energy disables gravity. Because of this dependence, as the universe cools, the tissue of spacetime goes through a very special condition when the fields (gravity/anti-gravity due to the dark energy) become balanced.

SSR equations reveal that the point of equilibrium between gravity and anti-gravity fields occurs in the following two special situations, namely when the particle speed is $V_0=\sqrt{cV}\sim 10^{-3}m/s$\cite{Rodrigo}\cite{tachyon} with respect to the absolute referential frame of SSR so-called ultra-referential $S_V$, which is related to the cosmological vacuum due to the minimum speed $V$, or when the BCR temperature is too close to a minimum temperature $T_{min}\sim 10^{-12}K$ so-called horizon temperature. 

In order to understand better the meaning of the speed $v_0$, let us first write the energy of a particle in SSR, as follows: 
 
\begin{equation}
 E=m_{0}c^{2}\frac{dt}{d\tau}=m_{0}c^{2}\frac{\sqrt{1-\frac{V^2}{v^2}}}{\sqrt{1-\frac{v^2}{c^2}}}, 
\end{equation}

from where we get
 
\begin{equation}
\Psi=\Psi(v)=\frac{dt}{d\tau}=\frac{\sqrt{1-\frac{V^2}{v^2}}}{\sqrt{1-\frac{v^2}{c^2}}}. 
\end{equation}

\begin{figure}
\includegraphics[scale=0.80]{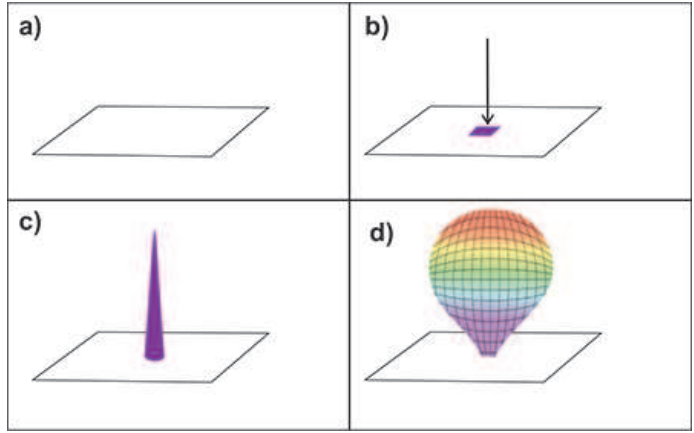}
\caption{A Newtonian or flat space works like an uncreated primordial universe. For a certain reason, this ``serene lake" or the primordial vacuum pre-existing to the creation (with null curvature, i.e., a flat space) is in the eminence of being disturbed. An infinite negative curvature arises generating a vacuum with a very strong anti-gravity which creates a high peak at the Planck scale. The temperature that increases drastically leads to the emergence of an inflationary bubble that will generate our universe. Probably, such primordial (trans-Planckian) vacuum or {\it nothing} is a source of many other universes, each of one with its own universal constants.} 
\end{figure}

So we write the energy $E$ as follows: 
  
\begin{equation}
E=m_{0}c^{2}\Psi,
\end{equation}
so that $v_0=\sqrt{cV}$ represents the speed that leads to the proper energy of the particle,
namely $E_0=m_0c^2$ as $\Psi(v_0)=1$. Such proper energy in SSR (for $v=v_0$) coincides exactly
with the rest energy (rest mass) in SR-theory. 

We define the potential of SSR, namely: 

\begin{equation}
 \phi=c^{2}\left[\Psi(v)-1\right], 
\end{equation}
such that we find $\Psi(v)\equiv(1+\phi/c^2)$, where $-c^2\leq\phi<\infty$. 
 
Let us show the scalar field $\phi$ connected to the vacuum represented by $S_{V}$. We should 
realize that such a field is a strongly repulsive scalar pontential, so that $\phi<<0$. This is a non-classical aspect of gravity for much lower energies given in the limit of vacuum, i.e.,
for $v\approx V$. Thus we write the following approximation by neglecting the Lorentz factor 
$\gamma$ (higher energies or $v>>v_0$), as follows: 

\begin{equation}
E\approx{m_{0}}c^{2}\Theta^{-\frac{1}{2}}\approx{m_{0}}c^{2}\left(1+\frac{\phi}{c^2}\right),
\end{equation}
where $\Theta^{-1/2}=\theta=\sqrt{1-V^2/v^2}$. 

In the limit $v\rightarrow V$, this implies $E\rightarrow 0$, which corresponds to 
$\phi\rightarrow-c^2$, i.e., this is the lowest potential for representing the most repulsive potential related to the vacuum $S_V$. So we find $\phi(V)=-c^2$ for representing the most fundamental vacuum potential. We have interpreted this result\cite{N2016} for describing an exotic particle (an infinitely massive boson) of vacuum that would escape from a 
maximum anti-gravity $\phi=-c^2$\cite{N2016}\cite{N2015}\cite{N2008}, but any finite massive particle could escape from such strongest anti-gravity working like an anti-gravitational horizon connected to a de Sitter (dS) horizon associated with a sphere of dark energy\cite{Rodrigo}. So it is interesting to note that the virtual particle of vacumm in $S_V$ has infinite mass since $\Theta^{-1}(v=V)=0$, thus being the counterpart of the photon, which is a massless particle as $\gamma(v=c)=\infty$. 

Under the conditions of gravity/anti-gravity equilibrium ($v=v_0$ and $\phi=0$) (Fig.7), the spacetime becomes Euclidean and exhibits momentarily (in a cosmological time) the characteristics of the Newtonian (flat) space.

In order to find the universe in this condition of equilibrium between gravity and antigravity (Fig.7), we consider the following reasoning: if a pendulum were at rest upon a first observation and caught in motion on a second observation, it is concluded that the state of rest was broken by the action of a cause represented by an {\it information}. Such performance would have occurred at any moment between the two peeps.

As a reminder, it should be noted that, as the universe cools down, it goes through a temperature $T_0$ when spacetime becomes flat (Newtonian). So, a particle with speed $v_0$ with respect to the ultra-referential $S_V$ also travels in a flat or Euclidean spacetime. But below $T_0$ (for global effect that involves all quanta in the universe) or below $v_0$ (for local effect or the case of a single particle), the spacetime takes on the dS-geometry\cite{Rodrigo}, i.e., the dark energy governs the universe with a positive cosmological constant. On the other hand, above such parameters ($v>v_0$ and $T>T_0$), gravity predominates when the spacetime displays the proper description of an AdS metric with a negative cosmological constant. 

\begin{figure}     
\centering
\includegraphics[scale=0.90]{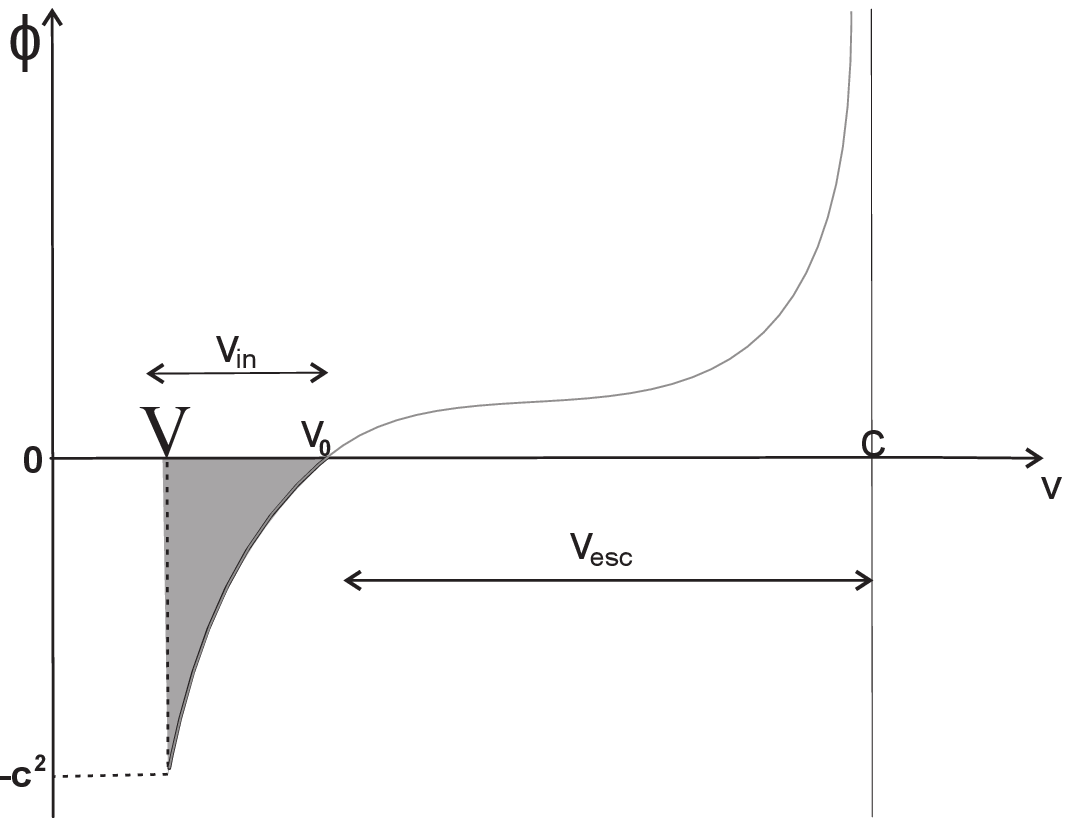}
\caption{This graph shows the general scalar potential $\phi(v)=c^{2}\left(\sqrt{\frac{1-\frac{V^2}{v^2}}{1-\frac{v^2}{c^2}}}-1\right)$ given in function of the velocity [Eq.(9)]. It shows the two phases, namely gravity (right side)/anti-gravity (left site), where the barrier at the right side represents the ultra-relativistic limit (speed of light $c$ with $\phi\rightarrow\infty$), and the barrier at the left side represents the quantum (anti-gravitational) limit only described by SSR (minimum speed $V$ with $\phi=-c^2$). The long intermediary region is the Newtonian regime ($V<<v<<c$), where occurs the phase transition just for $v=v_0=\sqrt{cV}$, with $\Psi(v_0)=1$.}
\label{Rotulo}
\end{figure}

The primordial vacuum or trans-Planckian vacuum pre-existing to creation is shown in Fig.6,a. It is like the surface of a serene lake given by a perfect flat spacetime (Newtonian scenario), and so it represents the 
uncreated universe, which provides the primordial vacuum for the creation itself. The sudden - but not random - action of an intervention set the primordial vacuum in motion, and due to its disruption, led it to focus on a singularity of a very high density and temperature, and thus to the big bang followed by the inflation due to a highly repulsive dark energy that allowed the emergence of spacetime. Hence, all the necessary conditions for the big bang were met. In no way we can say that this was an effect without cause, according to the cosmology of the Symmetrical Relativity, as we will show later. 

\section{Yale's experiment versus spontaneous transition: an interface with the cosmology of the spontaneous creation}

Superconducting material (e.g: niobium) is used to simulate quantum transitions in the artificial atom for the generation of qubits. The enabling switch is provided by the formation of Cooper pairs - at the required temperature - each time a magnetic field is triggered. The unpaired electrons transition to the pairing state when they pass to the superconducting regime.

The online edition of Nature on June 03 (2019)\cite{Yale} published an article dealing with the quantum leap, monitored and reversed, and revealed that it is a continuous process. The speed of the transition is very high and, until then, was considered instantaneous. This was not the view defended by Erwin Schroedinger, Einstein, de-Broglie and Bohm. 

The experiment indicates how randomness is only apparent. Not only can the transition be induced, but it can also be reversed.

The researchers were able to detect a quantum leap that was about to happen, interrupt it halfway and return the system to its current state.

Wires of superconducting material have, like electrons in real atoms, discrete states of energy. They are employed in the qubit generation architecture of the quantum information infrastructure.

A single artificial atom has been prepared in an aluminum $3D$-cavity, where $E_1$ energy photons are produced to induce a transition and $E_2$ and $E_3$ energy photons are generated for the purpose of indirect monitoring to prevent system collapse. 

During the quantum leap, the system goes through the overlap of both states. But, as the leap progresses, the likelihood of its completion is greater. Through the technique called tomographic reconstruction, it was possible to perceive the gradual weight change of the weights related to each of the two states. This was the monitoring stage, followed by paralysis and intervention in the process.

The possibility of intervening to reverse the jump once it has been predicted - albeit through observation - indicates continuity and that it cannot be random.

We know that during a quantum transition, energy becomes matter and vice-versa. The production of virtual particle pairs from quantum vacuum is also a recognized fact. We can also cite the Casimir effect. Among other facts, the aforementioned were important so that considerations pertinent to the concept of randomness - which emerged from experiences in QM laboratories - were passed on to cosmology, in a very natural reasoning. But a new horizon opened up from the discovery of Yale experiment in Yale Quantum Institute.

Now it is equally natural and necessary to revise the point of view defended by the authors of the spontaneous creation, the same view favorable to the proposals that consider {\it chance} as the cause of everything.

The current quantum leap experiment help us in clarifying this important issue.

\subsection{Nullifying Information Elements (NIEs) of the Spontaneous Creation}

We recognize that there is an important convergence between the spontaneous creation's NIE's interlacing conception (Fig.8) and SSR's ideas about causalities that connect, through the temperature of the BCR and the present instant (``instantaneous'' time), all the particles in the universe with each other and with the cosmological vacuum. However, there is a crucial distinction between 
the cosmology of SSR and the spontaneous creation to be presented later. 

In short, while the cosmology of SSR plays the role of the Yale experiment, where the collapse of the wave function is controlled by the observer within a certain amplified interval of time, a spontaneous very rapid transition, out of the observer control plays the role of the spontaneous creation in the cosmological scenario. 

\begin{figure}
\includegraphics[scale=0.3]{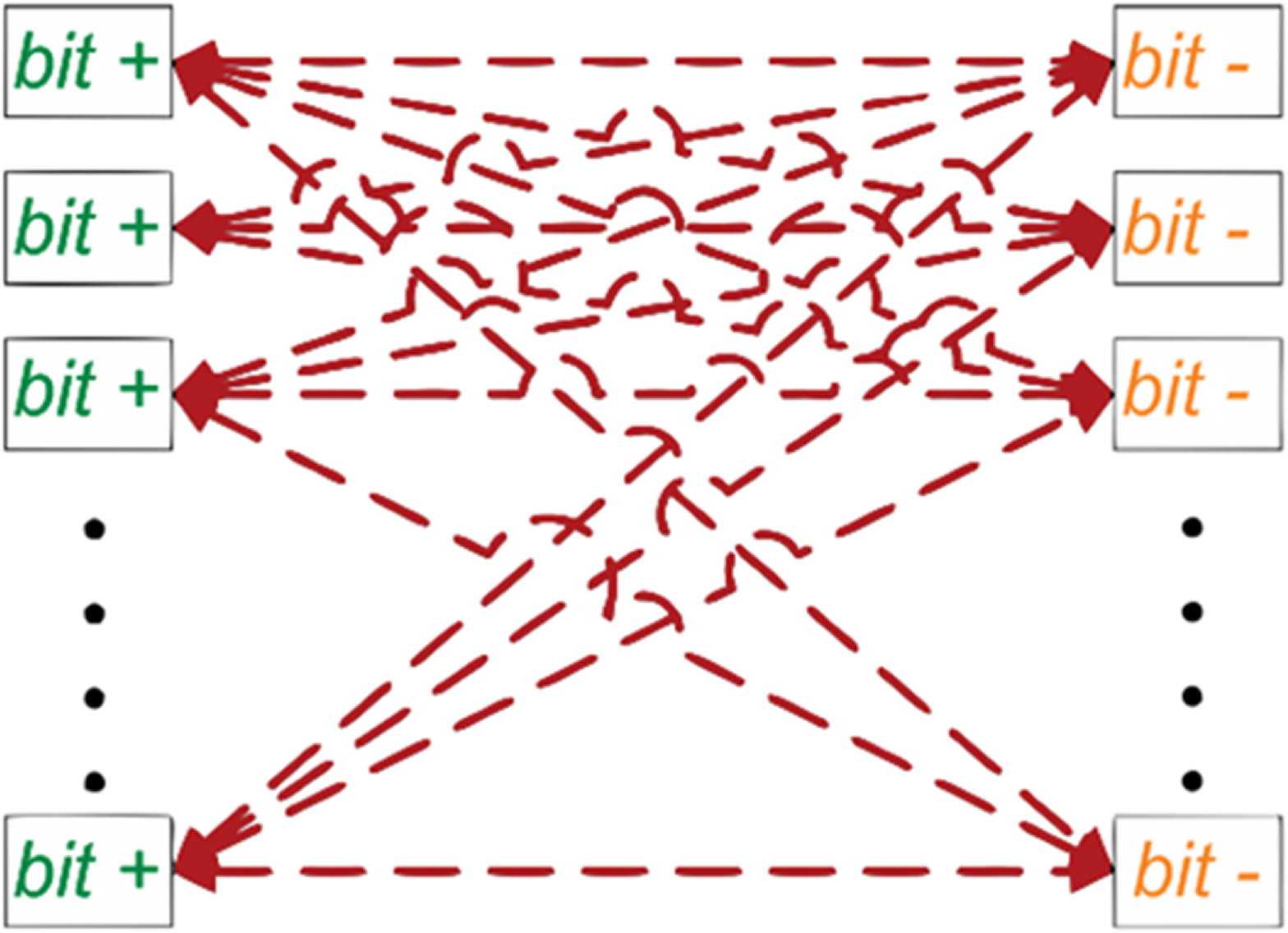}
\caption{In terms of information, {\it nothing} is equivalent to an infinite number of simultaneous NIEs\cite{CEN}. The dashedarrows symbolize possible co-existence relationships between bit $(+)$ and bit $(-)$ pairs.}
\end{figure}

\section{The quantum non-locality and the dark energy} 

A quantum transition is the shortest physical process in nature. Such transition has a duration
(proper time) practically null ($\Delta\tau=0$), as it occurs instantaneously with an infinite speed within the resolution of any experiments until the present time. Thus, at a first sight, it seems to violate the 2nd. postulate of Special Relativity (SR) that states that the speed of light $c$ is the maximum speed of a given signal between two points. However, in spite of the quantum effects by also including non-locality effects violate the causality in a classical spacetime like SR due to a kind of tachyonic (superluminal) ``signal'', it has been already shown by SSR\cite{tachyon} the existence of a non-local background field (vacuum energy) whose reference frame is associated with an invariant minimum speed for any particles, so that it corresponds to an infinitely massive boson, which leads to a reciprocal (informational) speed $\sigma c=c^2/V\sim 10^{30}$m/s extremely larger than the speed of light $c$, but still finite, as $\sigma=c/V\sim 10^{23}$\cite{tachyon}\cite{N2016}, which represents a dimensionless constant\cite{tachyon}\cite{N2016}. 

Thus, by considering that $c^{\prime}=\sigma c\approx 10^{31}$m/s involves an informational (superluminal) speed connected to the vacuum non-locality, and knowing the very small distance that separates the layers of the electrosphere of an atom in the order of about 
$10^{-12}$m, which is close to the order of the Compton wave length of the electron, i.e.,
$\lambda_C=h/m_ec\sim 10^{-12}$m, we can calculate the too small magnitude of transition
duration given in normal conditions (not so far from the room temperatures), namely 
$\Delta\tau_{transition}\sim\lambda_C/\sigma c=10^{-12}m/10^{31}m/s=10^{-43}s$. 

It is very important to notice that such tiny time interval 
$\Delta\tau_{trans}=\Delta\tau_{transition} (\sim 10^{-43}$s) represents exactly the well-known Planck time $t_P$ as being the smallest interval of time in nature. Thus we are led to conclude that the time of quantum transition for such normal conditions is non-null (but very small) in the spacetime with a minimum speed $V\propto L_P$\cite{N2016} (Planck length $\sim 10^{-35}$m), so that we can write

\begin{equation}
 T_P=\frac{L_P}{c}=\sqrt{\frac{G\hbar}{c^5}}\sim\Delta\tau_{trans}\sim \frac{\hbar V}{m_ec^2}
 \sim 10^{-43}s, 
 \end{equation}
with $V/c=\sigma^{-1}=\xi$\cite{N2016}. It is slready known that $L_P=\sqrt{G\hbar/c^3}$. 

Here we should stress that SSR-theory\cite{N2016} explains the apparent effect of non-locality as being of origin of a new 
relativistic effect that occurs in spacetime at lower energies, namely when the particle is too close to the invariant minimum 
speed $V$ (the preferred non-local reference frame $S_V$), so that there emerges a drastic increasing of the proper time (proper time dilation or $\Delta\tau\rightarrow\infty$ for $v\rightarrow V$), i.e., a drastic increasing of the so-called duration or proper time of the own particle, thus giving us the impression that it is in everywhere, which represents an effective 
superluminal signal $c'>c$ given by the improper referential in SSR; however, it is indeed an ilusion seen by the 
improper observer. Actually, such ilusion of violation of the speed of light limit comes from the dilation of the proper 
time as shown in the next subsection. 

\subsection{The concept of reciprocal velocity and its connection with the uncertainty principle: 
the apparent tachyonic signal and its connection with the vacuum energy}

As already discussed in a previous paper\cite{N2012}, SSR generates a kinematics of non-locality as also proposed in the emergent gravity theories\cite{acus2}. 

In order to see more clearly the aspect of non-locality of SSR due to the stretching of space when $v\approx V$, we should take intoaccount the idea of the so-called reciprocal velocity $v_{rec}$, which has been already well explored in a previous paper\cite{N2012}. Thus, here we will just reintroduce such idea in a more summarized way. To do that, let us use Eq.(7) and first multiply it by $c$ at both sides, and after by taking the squared result in order to obtain   

\begin{equation}
 \left(c^2-\frac{c^2V^2}{v^2}\right)(\Delta\tau)^2=(c^2-v^2)(\Delta t)^2, 
\end{equation}
where the right side of Eq.(12) related to the improper time $\Delta t$ provides the velocity $v$ of particle 
($\Delta t\rightarrow\infty$ for $v\rightarrow c$) and, on the other hand, the left side gives us the new information that shows that the proper time in SSR is not an invariant quantity as is in SR, so that the proper time goes to infinite ($\Delta\tau\rightarrow\infty$) when $v\rightarrow V$, which leads to a too large stretching of the proper space interval $c\Delta\tau$ in this limit of much lower speed, giving us the impression that the particle is well delocalized due its high internal (reciprocal) speed that is so-called reciprocal velocity that appears at the left side of the equation as being $v_{rec}=(cV)/v=v_0^2/v$. So we have $(c^2-v_{rec}^2)(\Delta\tau)^2=(c^2-v^2)(\Delta t)^2$. Now we can perceive that the reciprocal velocity $v_{rec}$ represents a kind of inverse of $v$ such that, when $v\rightarrow V$, we get $v_{rec}\rightarrow c$, i.e., the internal or reciprocal motion is close to $c$, thus leading to the new effect of {\it proper time dilation} associated to a delocalization that was shown as being an uncertainty on position\cite{N2012} within the scenario of spacetime in SSR. In this scenario, it was shown that the decreasing of momentum close to zero ($v\approx V$) leads to a delocalization of the particle, which is justified by the increasing of 
$v_{rec}\rightarrow c$ and the dilation of the proper 
time $\Delta\tau\rightarrow\infty$. As the uncertainty on position is $\Delta x=v_{rec}\Delta\tau=(v_0^2/v)\Delta\tau$\cite{N2012}, 
for $v\rightarrow V$, we find $\Delta x=c\Delta\tau\rightarrow\infty$. And, on the other hand, the large increasing of momentum for $v\rightarrow c$ 
leads to the well-known dilation of the improper time $\Delta t$ (right side of Eq.(12)), so that we find 
$\Delta\tau<<\Delta t$ and
the minimum reciprocal velocity $v_{rec}\rightarrow v_0^2/c=V$, which provides a small uncertainty on position, since 
$\Delta x=V\Delta\tau$.

Therefore Eq.(12) or Eq.(7) can be rewritten in the following way: 

\begin{equation}
\Delta\tau\sqrt{1-\frac{v^2_{rec}}{c^{2}}}=\Delta t\sqrt{1-\frac{v^{2}}{c^{2}}},
\end{equation}
where we find $v^2_{rec}/c^2=V^2/v^2$. We have $V<v<c$ and $V<v_{rec}<c$, where $V$ is the reciprocal of $c$ and vice versa.  
 
We claim that the concept of reciprocal velocity\cite{N2012} emerges in order to understand the apparent superluminal effects 
of tachyons as being the result of a new causal structure of spacetime with the presence of dark energy by representing $V$, so that the so-called tachyonic inflation is now explained as being an effect of
a drastic proper time dilation and large stretching of the space close to the vacuum regime governed by the phase of anti-gravity (Fig.6), thus leading to an inflation, i.e., a too rapid expansion of the space. Hence, according to the idea of reciprocal velocity, when we are close to $V$ ($S_V$), we find $v_{rec}(v\approx V)\Delta\tau=(v_0^2/V)(\Delta\tau)=(v_0^2/V)(\Delta t)\theta^{-1}(v\approx V)=[c\theta^{-1}(v\approx V)]
\Delta t=c^{\prime}\Delta t\rightarrow\infty$, giving us the impression of a superluminal speed $c^{\prime}=c\theta^{-1}>>c$, since we now
consider that the dilation factor $\theta^{-1}=(1-V^2/v^2)^{-1/2}$ is acting on $c$ instead of acting on $\Delta t$, but the result
$c^{\prime}=c\theta^{-1}$ is merely apparent, working like a tachyon. Actually SSR shows that occurs a stretching of space.  

In other words, in sum we say that the tachyonic inflation ($v=c^{\prime}>>c$) just mimics the new spacetime effects of SSR close to the invariant minimum speed $V$. 

\subsection{Tachyonic fluids as dark energy within the Machian scenario of SSR}

The relativistic Lagrangian of a free particle is 

\begin{equation}
 \mathcal{L}=-m_{0}c^{2}\gamma^{-1}=-m_{0}c^{2}\sqrt{1-\frac{v^2}{c^2}}, 
\end{equation}
but, if the particle suffers the effects of a conservative force, which is independent of speed, we have 
$\mathcal{L}=-m_{0}c^{2}\sqrt{1-\beta^{2}}-U$, where $U=U(r)$ is a potential depending on position. If $L$ is a function that does not depend on time, so there is a constant of motion $h$, namely: 

\begin{equation}
 h=\dot q_{j} p_{j}-\mathcal{L}=\frac{m_0v_{j}v_{j}}{\sqrt{1-\frac{v^2}{c^2}}}+m_{0}c^{2}\sqrt{1-\frac{v^2}{c^2}}+U,  
\end{equation}
which leads to 
\begin{equation}
 h=\frac{m_{0}c^2}{\sqrt{1-\frac{v^2}{c^2}}}+U=E, 
\end{equation}
where $h$ is the total energy. If $U=0$, we get $E=\gamma m_0c^2$, which is the energy of a free particle,  playing the role of the constant of motion. For a single particle, we write $v^2=v_jv_j$. 

Now, by using an analogous procedure to that one for obtaining the relativistic Lagrangian, we can get
the Lagrangian of SSR. 

Let us first obtain the Lagrangian of a free particle. Thus by considering $h=E=m_0c^2\Psi$ as the constant 
of motion for a free particle in SSR and knowing that $p=m_0v\Psi$ is the momentum, we obtain

\begin{equation}
 E=m_{0}c^{2}\sqrt{\frac{1-\frac{V^2}{v^2}}{1-\frac{v^2}{c^2}}}=h=m_{0}v^{2}\sqrt{\frac{1-\frac{V^2}{v^2}}
 {1-\frac{v^2}{c^2}}}-\mathcal{L}, 
\end{equation}
where $v^2=v_jv_j$. $\mathcal{L}$\cite{N2016} is the Lagrangian of a free particle in SSR.

From Eq.(17), we get

\begin{equation}
 \mathcal{L}=-m_{0}c^{2}\theta\sqrt{1-\frac{v^2}{c^2}}=-m_{0}c^{2}\sqrt{\left(1-\frac{V^2}{v^2}\right)\left(1-\frac{v^2}{c^2}\right)}.
\end{equation} 
If we make $V\rightarrow 0$ (or $v>>V$) in Eq.(18), we recover the relativistic Lagrangian of a free particle given in Eq.(14). 

In the presence of a potential $U=U(r)$, we write

\begin{equation}
\mathcal{L}=-m_{0}c^{2}\sqrt{\left(1-\frac{V^2}{v^2}\right)\left(1-\frac{v^2}{c^2}\right)}-U. 
\end{equation}

Now let us consider the case of a particle in the presence of an electromagnetic field. In this case\cite{N2016}, we have a non-conservative potential that depends on position and speed. So we obtain

\begin{equation}
\mathcal{L}=-m_{0}c^{2}\sqrt{\left(1-\frac{V^2}{v^2}\right)\left(1-\frac{v^2}{c^2}\right)}-q\Phi+\frac{q}{c}\vec{A}\cdot\vec{v}. 
\end{equation}

If there is no charge, i.e., $q=0$, we recover Eq.(18) that represents the Lagrangian of a free particle
in SSR. 

We intend to show that the Lagrangian in Eq.(18) generates non-local effects that remind the Machian effects when the speed $v$ is too close to the minimum speed $V$ connected to the ultra-referential $S_V$, thus leading to a strong coupling of the particle with vacuum, so that its mass becomes strongly dressed by the vacuum energy. Thus, as the particle approaches more and more to the vacuum regime $S_V$, 
its coupling with vacuum increases drastically and then its dressed mass increases a lot, such that the particle becomes extremely heavy due to the global effect of vacuum energy. Such effect so-called dressed mass\cite{N2015} has a Machian origin
within a quantum scenario of a gravitational vacuum. The dressed mass $m_{dress}$ of a particle shown in the approximation of vacuum regime 
given by the mimimum speed $V$ was investigated before\cite{N2015}, where we have found 
$m_{dress}\approx m_0/\sqrt{1-V^2/v^2}$.

On the other hand, SSR predicts that the bare (relativistic)  mass $m\approx m_0\sqrt{1-V^2/v^2}$
goes to zero when $v\rightarrow V$, so that the ``particle'' loses its identity close to the vacuum-$S_V$ ($m\approx 0$) and so it becomes completely delocalized, i.e., a uncertainty on position as we get
$c\Delta\tau\rightarrow\infty$\cite{N2012}, since it merges with vacuum as a whole, thus
leading to $m_{dress}\rightarrow\infty$. 

The impossibility of reaching the mimimum speed $V$ is due to the increase of the dressed mass that goes to infinite for $v\rightarrow V$ and thus prevents the increasing of its decelaration in such a way that 
a too delocalized and so heavy particle cannot reach the minimum speed.

\subsection{The Dirac-Born-Infeld Lagrangian and its connection with SSR-Lagrangian: the apparent superluminal effect $\sigma c$ and its connection with the non-locality}
 
The Dirac-Born-Infeld (DBI) Lagrangian\cite{li},\cite{rola, rola2, rola3, dark} that presents identical characteristics of a tachyonic (superluminal) fluid is of the type $L=\mathcal{V}(\phi)\sqrt{1-(\partial_{t}\phi)^2}$, where $\phi$ is a scalar field and $\mathcal V(\phi)$ is a tachyonic potential. The potential 
$\mathcal{V}(\phi)$ is known as the Dirac-Born-Infeld (DBI) potential. We should perceive that such a Lagrangian is similar to SSR-Lagrangian of a free particle given in Eq.(18), since we have already considered that the speed $v$ plays the role of a scalar field related to the gravitational potential $\phi$ of SSR.

Here we intend to go deeper in exploring the connection between DBI Lagrangian and SSR-Lagrangian by reinterpreting the tachyonic 
inflation (superluminal effects) within a new Machian scenario of graviting vacuum generated by spacetime of SSR when the concept 
of dressed mass naturally emerges from the scenario of DBI when compared with the scenario of SSR. To do that, by comparing 
DBI Lagrangian with SSR Lagrangian given in Eq.(18), we can identify the DBI potential $\mathcal{V}$ as being the term of SSR 
Lagrangian that provides the negative potential $-c^2$ multiplied by the factor $\theta(v)$, namely: 

\begin{equation}
 \mathcal{V}=-c^2\theta=-c^{2}\sqrt{1-\frac{V^2}{v^2}}\equiv-c^{2}\left(1+\frac{\phi}{c^2}\right),  
\end{equation}
where we have $v\approx V$ for $\phi\approx -c^2$. The DBI potential is within the SSR-scenario. 

The reason of considering $\theta$ for obtaining $\mathcal{V}$ is due to the fact that just $\theta$ allows to get corrections close
to vacuum ($v\approx V$), since vacuum is responsible for an inflation governed by anti-gravity.

As $\mathcal{V}$ in Eq.(21) is a kind of DBI potential within SSR-scenario, we can rewrite SSR Lagrangian in Eq.(18) in the following way: 

\begin{equation}
 \mathcal{L}=m_{0}\mathcal V\sqrt{1-\frac{v^2}{c^2}}, 
\end{equation}
where $\mathcal V=\mathcal V(v)\equiv\mathcal V(\phi)$ according to Eq.(21). As we already know that the speed $v$ also plays the
role of a scalar field, let us first consider the DBI potential as a function $\mathcal V=\mathcal V(v)$ for our proposal in 
finding the Machian scenario that should emerge from Eq.(21). And, since such Machian scenario is within the context of gravitational vacuum for $v\approx V$, let us neglect the Lorentz factor in Eq.(22) by making the following approximation:  

\begin{equation}
\mathcal{L}\approx m_{0}\mathcal V, 
\end{equation}
where $\mathcal V=\mathcal V(v)=-c^2\theta(v)$. 

Let us obtain the equation of motion from the Lagrangian given in Eq.(23). Therefore, we should use Euler-Lagrange equation. As $\mathcal L$ just has dependence on speed $v$, we obtain

\begin{equation}
\frac{d}{dt}\left(\frac{\partial\mathcal L}{\partial v}\right)= m_0\frac{d}{dt}\left(\frac{\partial\mathcal V}{\partial v}\right)=0, 
\end{equation}
from where we get the generalized momentum $\mathcal P$ of SSR given at lower energies, namely: 

\begin{equation}
\mathcal P= m_0\frac{\partial\mathcal V}{\partial v}=-m_0c^2\frac{d\theta}{dv}, 
\end{equation}
where $\theta=\theta(v)=\sqrt{1-V^2/v^2}$. So, by introducing $\theta(v)$ into Eq.(25), we find  

\begin{equation}
\mathcal P=\mathcal P_{machian}=\frac{m_0}{\sqrt{1-\frac{V^2}{v^2}}}\left(\frac{c^2V^2}{v^3}\right),  
\end{equation}
which can be written in the following way: 

\begin{equation}
\mathcal P_{machian}=m_{dress}\left(\frac{v_0^4}{v^3}\right),  
\end{equation}
where $v_0=\sqrt{cV}$ and we find that the generalized momentum $\mathcal P$ depends directly on the dressed mass 
$m_{dress}=m_0/\sqrt{1-V^2/v^2}$\cite{N2015} as expected from DBI theory within SSR scenario. Therefore, we conclude that 
the generalized momentum $\mathcal P_{machian}$ is interpreted as being of Machian origin within a scenario of graviting vacuum, since
such Machian momentum increases drastically to infinite when $v\rightarrow V$ due to a very strong coupling of the particle $m_0$ with 
graviting vacuum at the universal reference frame $S_V$, thus leading to an effective mass dressed by vacuum, $m_{dress}$ being 
much larger than the bare mass $m_0$ given for $v=v_0$, i.e., we get $m_{dress}(v\approx V)>>m_0(v=v_0)$. So, we 
can finally conclude that a particle too close to the vacuum $S_V$ is extremely heavy and carries out a very high generalized
momentum that can be obtained from Eq.(27) by making $v\approx V$, so that we can write 

\begin{equation}
\mathcal P_{(v\approx V)}\approx m_{dress}\left(\frac{v_0^4}{V^3}\right),  
\end{equation}

or we can alternatively write this high generalized momentum of a ultra-heavy virtual particle of vacuum, as follows: 

\begin{equation}
\mathcal P_{vac}=m_{dress}(\sigma c),  
\end{equation}
where we have $c^{\prime}=\sigma c=(c/V)c=\xi^{-1}c\sim 10^{31}m/s$, with $\sigma=\xi^{-1}=c/V\sim 10^{23}$ and $c\sim 10^{8}m/s$. 

According to Eq.(29), we realize that a particle which is too close to the vacuum $S_V$ merges with the own vacuum fluid having a dressed mass that ``travels" as if it were a tachyon moving with the highest superluminal speed 
$c^{\prime}=\sigma c\sim 10^{31}m/s$ associated with the idea of non-locality of the ultra-referential $S_V$, so that, if we admit the Hubble radius $r_H(\sim 10^{26}m)$ for the universe, such a tachyonic fluid with speed $\sigma c(\sim 10^{31}m/s)$ would propagate across the whole universe during only about $r_H/\sigma c\sim 10^{-5}s$. However, within the SSR scenario, it is important to stress that such superluminal effects are apparent in the sense that such effects are in fact generated by a large stretching of spacetime when $v$ is too close to $V$ (vacuum regime), so that we find $\Delta x^{\prime}=c(\Delta\tau)\approx c(\Delta t/\sqrt{1-V^2/v^2})\rightarrow\infty$ (Eq.(7) for $v\approx V$), giving us the impression that we have a superluminal signal $c^{\prime}=c/\sqrt{1-V^2/v^2}$.

Finally, Eq.(29) permits us to obtain the vacuum energy associated with its non-locality due to the large dimensionless scale 
factor $\sigma=c/V\sim 10^{23}$\cite{N2016}, namely: 
    
\begin{equation}
 E_{vac}=\mathcal P(\sigma c)=m_{dress}\sigma^2 c^2,
\end{equation}
where $\sigma c\sim 10^{31}m/s$ is the apparent superluminal effect that leads to the spatial correlation between two points 
separeted by a so long distance in the space. This (quantum) non-locality effect due to a too rapid propagation of ``signal''
({\it information}) by means of the vacuum energy is the cause behind the Machian effects, in spite of the Mach principle is merely a classical principle in the sense that it takes into account a {\it distance action} with an infinite speed across the empty space (classical vacuum), whereas the so-called {\it quantum Mach effect} due to a finite $\sigma c$ 
-but a too high speed- across the quantum vacuum leads to the most fundamental vacuum energy-dark mass relation given in Eq.(30). It is important to notice that Eq.(30) has the same form of the famous mass-energy relation of Einstein $E=mc^2$, however $m$ is 
for matter, whereas $m_{dress}$ is an effective mass related to the dark energy of vacuum, so that we can alternatively write 
Eq.(30)\cite{tachyon} as $E_{vac.}=m_{vac.}\sigma^2 c^2$, where $m_{vac}$ represents a certain portion of vacuum dark mass. 

It is surprising to realize that vacuum [Eq.(30)] stores much more energy than matter, although the vacuum energy is diluted throughout the space. Therefore, Eq.(30) is a fundamental implication of SSR\cite{N2016}, as well as $E=mc^2$ is a fundamental implication of Special Relativity (SR). 

\subsection{The invisible radius of the universe: the dark radius} 

By considering that, for each quantum time (Planck time $T_P\sim 10^{-43}s$) with the superluminal speed 
$\sigma c\sim 10^{31}m/s$, there emerged a primordial atom with radius of about $\sigma c T_P\sim 
10^{31}m/s\times 10^{-43}s\sim 10^{-12}m$, which represents exactly the order of magnitude of the electron Compton wavelength. Thus, by taking into account the age of the universe (about $15$ billions of years) with such a tachyonic (superluminal) speed $\sigma c$, we can calculate its total (invisible) extension, which is indeed much larger than the visible (Hubble) radius $R_H\sim 10^{26}m$. 

Let us now compare such invisible radius of the dark universe with the Hubble radius of the visible universe. By means of a simple calculation, we will obtain the invisible radius of the dark universe, which is equal to
$(\sigma c)(R_H/c)\sim 10^{23}\times 10^{26}m\sim 10^{49}m$. 

As the Hubble radius is $R_H\sim 10^{26}m$, we conclude that the true radius of the dark (invisible) universe is $23$ order of magnitude greater than the Hubble (visible) radius.

\section{The Planck time as the quantum of time for representing a quantum transition, the Yale experiment, the spacetime correlations and their interface with cosmology of the symmetrical relativity} 

When we investigate the subatomic world, although spatial limitation also obviously exists, what really imposes on us boundaries is time, since processes become increasingly ephemeral.

There will be a limit and it will certainly have to be the time of the order of $T_P\sim 10^{-43}s$ to be obtained (by approximation) in the laboratory. 

Physicists conducting Yale experiment with a transition in virtually minimal time will have achieved the ability to act on the big bang time. Thus, by operating at the time of a transition is in any way similar to act at the initial instant of the big bang.

From the cosmological point of view, it is a very intuitive matter that the start of big bang occurred in an extremely brief but absolutely continuous period of time. However, it has been established that $T_P\sim 10^{-43}s$ is the quantum of time associated with the extremely small duration of beginning of the big bang. 

But many difficulties were encountered in trying to understand the concept of quantum leap and make it compatible with the idea of the continuous transition of the primordial atom. In fact, because there is no jump, the transition is really continuous. This is why, although the perspective of general theory of gravitation is obtained with the view of continuity of time. This condition has never reconciled with the quantum scale of leap in the energy transition of an atomic-nuclear system, according to experiments of QM. 

As there is a minimum speed $V$ and a minimum temperature $T_{min}$\cite{N2018}, there is also a minimum time limit of the order of Planck time $T_P\sim 10^{-43}s$, as we have announced above, when time itself tends to be too small or too close to zero, considering a quantization of time for obtaining a quantum of time in the order of $T_P$. 

The time equation given in Eq.(7) shows that the proper time $\Delta\tau$ tends to infinite when the speed $v$ is too close to the minimum speed $V$. In a previous work\cite{N2018}, it was shown that 
the minimum speed $V$ is equivalent to a minimum temperature $T_{min}=M_PV^2/K_B(\sim 10^{-12}K)$ in the termodynamics context, 
so that the proper time of an atomic clock immersed in a ultra-cold gas also tends to infinite when its temperature approaches
to such minimum temperature, i.e., in the limit of lower temperatures, we have obtained 
$\Delta\tau\approx\Delta t/\sqrt{1-T_{min}/T}$\cite{N2018}, where $\Delta t$ is the improper time measured in the laboratory.

Now, if we admit that the improper time ($\Delta t$) measured by the clock of the observer in lab is quantized in so tiny quanta 
of time given by the Planck time (minimum time) $T_P$, then it is important to conclude that such a quantum of time could be
drastically increased inside the ultra-cold gas, so that the lower temperatures too close to $T_{mim}$ are able to lead to a 
infinite dilation of the minimum proper time represented by $T_P$ that increases to a certain dilated
quantum of time $T'_P$ by going to infinite for $T\rightarrow T_{min}$, i.e., we are able to write the dilation of the minimum duration or quantum of proper time in function of the temperature $T$ for a
ultra-cold system, namely: 

\begin{equation}
T'_P (T)\approx\frac{T_P}{\sqrt{1-\frac{T_{min}}{T}}}, 
\end{equation}
where $T_P=\sqrt{G\hbar/c^5}\sim 10^{-43}s$ is the minimum time (quantum of time or Planck time).

The great novelty is that Eq.(31) indicates that too low temperatures are able to change drastically the 
duration of such quantum of time $T_P$ that provides an infinitesimal jump between a given instant $t_{n}$ and its successive instant $t_{n+1}$. It would be as if we had $t_{n+1}=t_{n}+T_P$ in the normal 
conditions of higher temperatures far from the minimum temperature $T_{min.}(\sim 10^{-12}K)$, i.e., 
$T>>T_{min.}$. However, according to Eq.(31), it is very important to stress that the temperature by representing essentially a vibration is the fundamental physical greatness, which is deeply linked with the concept of duration, in the sense that the stretching of the duration (proper time) has origin in 
the own increasing of the {\it quantum jump of time} $T'_P (T)$ between $t_{n}$ and $t_{n+1}$ when the 
system becomes ultra-cold, i.e., when $T\approx T_{min.}$. So the successive instant $t_{n+1}$ can be even
more distant from its imediately previous instant $t_n$ when the temperature decreases, so that we write 

\begin{equation}
 t_{n+1}=t_n +\frac{T_P}{\sqrt{1-\frac{T_{min}}{T}}}, 
\end{equation}
where Eq.(32) for $T>>T_{min}$ gives us the idea that time itself is almost continuous, since we may think that the current universe is analogous to a strobe that vibrates with such a so high frequency in the order of $\nu_P=T_P^{-1}\sim10^{43}$Hz 
(Planck frequency), which gives us the perception of time continuity. However, the novelty of Eq.(32) is that it shows that the tiny temporal leap ($T_P$) extends to infinity when the temperature of the universe tends to the minimum temperature, i.e., 
$T'_P\rightarrow\infty$ when $T\rightarrow T_{min}$. At a first sight, this means that a ultra-cold universe works like a strobe 
with a quasi-zero frequency, which will occur in a very distant future within the cosmological scenario. 

The most important interpretation of the temporal leap ($T'_P$) is that it provides a kind of 
correlation persistence between two spacetime events, so that the ultra-cold systems (e.g: the universe)
leads to a very large correlation persistence, which means that all the particles of a given system 
become strongly correlated beteween themselves as if the speed of light were infinite. In the cosmological
scenario, this will also occur at the end of the universe when its temperature
$T\rightarrow T_{min}$ [see Eq.(5)]\cite{TVSL}. 

We must stress that such a so strong (large) correlation persistence between two spacetime events gives us the idea of a very high synchronization between two events. Such a very high synchronization also occurred in the early universe when its temperature was too high close to the Planck temperature [Eq.(5)]. 

The interesting aspect of such synchronization in the viewpoint of QM (e.g: Yale experiment) is that it provides a strong correlation between two successive measures in a certain
ultra-cold system (e.g: niobium used in the experiment), in such a way that the probability of a first measurement in lab (collapse of a wave function) at an instant $t_n$ is practically the same probability of
a successive measurement at the instant $t_{n+1}$. In other words, this means a strong correlation of 
probabilities, thus leading to a deterministic behavior of such ultra-cold system between both
measurements. So the very low temperatures allow the observer to be able to control the instants when the wave functions collapses occur as was done in the Yale experiment. 

Therefore, now it is important to note that the present spacetime theory (SSR) leads to the cosmology of the Symmetrical Relativity whose equations given by Eq.(5)\cite{TVSL} and the new Eq.(32) provide the cosmological foundations as an interesting analogy for understanding what happens in the laboratory with the ultra-cold systems with respect to be able to control the instants of wave-functions collapses. So, we are led to conclude that there is a subtle connection between the increasing of spacetime correlations associated with the strengthening of the implicit determinism and the amplification of the observer's capacity of controlling the instants of collapse of wave functions, by surpassing the quantum probabilities, which would be the foundation of the future quantum computation. 

In sum, the cosmology of the Symmetrical Relativity\cite{TVSL} allows us to build an interface of {\it spacetime correlation}/
{\it implicit determinism}/{\it information}, by providing a deeper understanding of the foundations of QM (in Bohm`s context) 
that leads to the Yale experiment. 

Basing on Eq.(32), such interface can be represented mathematically by a correlation $\mathcal C$ function between two any instants $t_{n+m}$ and $t_n$, with $m=1,2,3...$ defined by the following way: 

\begin{equation}
\mathcal C(t_{n+m},t_n)=e^{\left[1-\frac{(t_{n+m}-t_n)}{T'_P}\right]}, 
\end{equation}
where $T'_P=T_P/\sqrt{1-T_{min}/T}$.

As $T'_P$ represents the minimum temporal leap for a certain temperature $T$ given in the approximation of temperatures much lower than the Planck temperature, we must have $t_{n+m}-t_n\geq T'_P$, where $T'_P$ can also be understood as being the persistence time 
of correlation between two successive instants $t_n$ and $t_{n+1}$, such that $t_{n+1}-t_n=T'_P$ [Eq.(32)].

Therefore, we should realize that the correlation function $\mathcal C$ obeys the following interval, namely $0<\mathcal C(t_{n+m},t_n)\leq 1$. For $m=1$, the correlation is maximum, i.e., 
$\mathcal C(t_{n+1},t_n)=1$, since we get $t_{n+m}-t_n\equiv t_{n+1}-t_n=T'_P$ in Eq.(33). For $m=2$, the 
correlation decreases to $\mathcal C_2=e^{-1}\approx 0.36787944117$. For $m=3$, we find 
$\mathcal C_3=e^{-2}\approx 0.13533528323$. So, of course for $m\rightarrow\infty$, the temporal correlation 
is zero, i.e., $\mathcal C_{\infty}=0$. The interpretation for such a null temporal correlation is the fact that the probability of any event given in an infinitely distant time is zero, which means a complete indeterminism or an infinite uncertainty for the occurrence of any event. As the actual correlation 
persistence time is so tiny of the order of $T_P\sim 10^{-43}s$ for temperatures $T>>T_{min}$, Eq.(33) indicates that even for a later instant like for instance $10^{-20}s(>>10^{-43}s)$, the temporal correlation function $\mathcal C$ already decreases drastically close to zero, which means that even such a later instant very close to the previous one already becomes practically uncertain for any event that is anticipated. That is why even a very near future is already completely uncertain for us as the quantum of time (Planck time $T_P$) is extremely tiny $\sim 10^{-43}s$.

However, the great novelty is that Eq.(33) shows that the future could be absolutely predictable with 
$100\%$ of probability ($\mathcal C=1$) only if the temperature of the system tends to $T_{min}$, leading 
to $T'_P\rightarrow\infty$. So, in this ideal condition of the lowest vibration or the minimum speed $V$ related to the minimum temperature $T_{mim}$, i.e., $T_{min}=M_PV^2/K_B$\cite{N2018}, even for an infinitely distant 
future $t_{m+n}-t_n$ with $m=\infty$, the correlation function or the probability $\mathcal P=\mathcal C$ [Eq.(33)] to antecipate any event would be $\mathcal P=1$, i.e., $100\%$ of certainty to predict any future event. This means that an ideal or 
{\it fictional observer} would be exactly at the preferred referential (ultra-referential) $S_V$ associated with the invariant 
mininum speed of SSR\cite{N2016}, which is responsible for the cosmological constant\cite{N2016}\cite{N2015}.

Such fictional observer is absolute in the sense that the all the points of the spacetime become infinitely correlated between 
themvelves. For this reason, the interface 
{\it spacetime correlation}/{\it implicit determinism}/{\it information} would be absolutely evident for such absolute {\it observer} as there would be an infinite {\it information}, working like an absolute implicit determinism (reminding Bohm's theory of the implicit order) behind all the (local) observers who are unable to determine the future as they are far from the background frame 
$S_V$, working like an an absolute implicit determinism behind all the (local) observers who are unable to determine the future as they are far from such background frame. This brings us to that idea of Newtonian absolute space, however within a modern framework given by SSR and its extension to the cosmology of the Symmetrical Relativity. Therefore, such cosmology shows us that the conception of Newton linked to an absolute space- now connected to a minimum speed and minimum temperature- makes sense within this new cosmological scenario that establishes the conception of the absolute implicit determinism, thus leading to an infinite spatial correlation [$c(T_{min})=\infty$ in Eq.(5)] and also a maximum temporal correlation of all the instants, i.e., 
$\mathcal P=\mathcal C(t_{n+m},t_n;T_{min})=1$ for $m=1,2,3......\infty$ [Eq.(33)]. 

Therefore, we should conclude that such a so intense {\it information} by means of a too strong spacetime correlation ($\mathcal C=\mathcal P=1$) is a natural extension of the present theory to an extreme condition, according to Eq.(5) and Eq.(33) when $T=T_{min}$. In view of this, SSR and the cosmology of the Symmetrical Relativity are upholding the idea of very strong 
implicit correlations behind all local spacetime events.

In the limit of $T_{min}$\cite{N2018} or even for $v\rightarrow V$ in Eq.(7)\cite{N2016}\cite{N2018}, it 
was already shown that the proper interval of time $\Delta\tau\rightarrow\infty$ suffers a large dilation, 
which means that occurs an infinite dilation of the proper time (the internal time of the particle) with respect to the improper one $\Delta t$ given to the observer in the laboratory. So, according to the concept of duration, which is directly related to the own physical internal time (atomic clock) or proper time of the object (e.g: the particle), we say that, when too close to $T_{min}$ or even $V$, the so-called objective duration or the proper time dilates to infinite (see Eq.(7)). However, we must stress that an infinite duration cannot be understood as a duration itself, since the own conception of duration itself must be essentially finite. However, we need a deeper physical understanding for such a finitude of time.

Since the origin of time is its quantization based on a finite quantum of time $T'_P$ [Eq.(32)] as being a minimum time with temperature dependende for representing a temporal leap or a persistence correlation time between a given instant $t_n$ and its successive instant $t_{n+1}$, we should realize that the nature of time given by own the concept of finitude denominated as duration makes sense only if it is quantized or made of the sum of an enormous amount of successive time packages so-called
{\it temporal quanta}, giving us the idea of a strobe with frequency 
$\nu_P(T)=(T'_P)^{-1}$. Thus, the non-zero finite frequency of this strobe that gives us the impression of the continuity of time is what guarantees the existence of duration, i.e., the physical 
condition for the existence of duration is the following inequality that must be obeyed: 
$0<\nu_P(T)<\infty$. But, if such inequality is not obeyed for extreme cases as, for instance, 
$\nu_P(T=T_{min})=0$ as considered before, the correlation persistence time $T'_P$ goes to infinite, i.e., the conception of duration does not make sense, as it must be finite itself. In other words, we say that the 
interval of proper time $\Delta\tau=\infty$ (for $T=T_{min}$ or even $v=V$ in Eq.(7)) is inconceivable.  

\subsection{Further considerations about the conception of the proper time in SSR} 

Time is the link between the physical and the ideal (extreme) condition given by Eq.(33) 
when we make $T=T_{min}$, thus leading to $T'_P=\infty$, i.e., the annulment of the conception of 
duration (proper time) as it must be always finite itself. 

Space dilation and improper time contraction occur locally when the particle speed approaches to $V$ [Eq.(7)], or globally when the BCR temperature approaches to $T_{min}$ as determined by the SSR equations for very low energies, which lead to the so-called 
{\it delocalization} of the particle\cite{uncertainty}, when it is literally everywhere. It is not just, as in the case proposed in the Schroedinger Wave Interpretation (IOS) of chances of being in several overlapping states at a single moment to collapse into only one due to measurement. Actually, it is the physical dilation of space itself, which is an exclusive effect of SSR spacetime, along the lines of Relativity Theory. But let us remember that Einstein sad exclusively about the contraction of space and the dilation of time only under high energy conditions, i.e., only if $v\rightarrow c$. 

The low energy conditions (very low temperatures) were essential for the work at the Yale Quantum Institute\cite{Yale}, as the experiment is performed in a ultra-high vacuum, as required by qubit ({\it information}) processing and employing superconductivity.

As we gain greater control of this experiment due to the increasing of spacetime correlations at lower temperatures as shown 
analogously in Eq.(33) within the new cosmological scenario, we will be pursuing the goal of overcoming the fragmentation of time (duration) to achieve the continuity of infinity. 

At once, we cannot have boundary conditions and infinity. Since all the processes are limited by duration, anyone can differ from any magnitude at all. The universe itself is finite. Space-time, energy of all fields, mass and cosmic existence itself are all finite because they are quantized, and the Planck time is a quantum of time. 

For this reason, the relativity theories- because of the mass, the energy of the fields and the spacetime present divergences in the limit of the speed of light - face classic incompatibilities with QM. Thus we have implicit infinities in the spacetime equations of relativity, and experimentally imposed quantization in QM formalism.

\section{Impossibility of spontaneous transition from trans-Planckian to Planckian regime in the big bang due to a very strong violation of the 2nd. law of thermodynamics in pre-big bang cosmology}

According to Eq.(33), the probability $\mathcal P(=\mathcal C)$ to occur any event in a distant future instant tends to 
zero, as we have considered the correlation persistence time $T'_P$. So, now by using the concept of Gibbs's entropy 
$\mathcal S=-K_B ln\mathcal P$ for representing Eq.(33) just for higher temperatures close to the big bang associated to the Planck time, such that $T'_P=T_P$, we obtain the following entropy for the early universe: 

\begin{equation}
\mathcal S=-K_B\left[1-\frac{(t_{n+m}-t_n)}{T_P}\right],  
\end{equation}
where $T_P(\sim 10^{-43}s)$ is the Planck time (a Planck unit or qubit) at the beginning of the universe, with $m=1,2,3...$ 
in Eq.(34).

At the beginning of the big bang, we notice that $m=1$ in Eq.(34), such that $t_{n+1}-t_n=T_P$, thus leading
to a null entropy, i.e., $\mathcal S=0$. This means that the universe emerged from a total order, which was leading to a bigger dis-order or an increased entropy ($\mathcal S\geq 0$) as it has expanded. 

But the great puzzle is: How did such a total order ($\mathcal S=0$) arise from a primordial (trans-Planckian) vacuum ({\it nothing}), which is investigated by the spontaneous creation ex nihilo (NIEs) by considering {\it nothing} as being effectively a soup of infinite bits $(+)$ and bits $(-)$ or an infinite sea of positive and negative virtual particles?

Based on the spontaneous creation, in the midst of an infinite soup of bits $(+)$ and bits $(-)$, that is to say, $50\%$ of $(+)$
and $50\%$ of $(-)$, i.e. {\it nothing}, several spontaneous processes of successive attempts with 
errors and hits were leading essentially to a high level of self-learning like a kind of Artificial 
Intelligence (AI), in such a way that the probability of breaking the balance (the symmetry) of bits $(+)$
and bits $(-)$ has increased by itself until reaching the condition of creation by means of the big bang,
when the order ($\mathcal S=0$) arose from such successive spontaneous attempts in the midst of chaos or 
{\it nothing} ($\mathcal S=\infty$) by {\it chance}. In other words, according to the computational view, we could say that the {\it primordial vacuum} made of a soup of $50\%$ of bits $(+)$ and $50\%$ of bits
$(-)$, whose sum leads to {\it nothing} represents only the hardware of a possible self programmable computer 
(the universe) to be created spontaneously or by the process of self-learning by starting from {\it nothing},
i.e., the so-called spontaneous creation\cite{CEN}. Thus, the software for such hardware would emerge spontaneously, i.e., an asymmetry ($+/-$) would spring out of {\it nothing} by {\it chance}, thus leading to the universe by means of the big bang when the entropy was $\mathcal S=0$.

On the other hand, the cosmology of the Symmetrical Relativity\cite{TVSL}, which is based on the conformal geometry of SSR-spacetime\cite{N2016}\cite{Rodrigo} with the implementation of the temperature of the universe, leads to exact quantifications of spacetime correlations given by Eq.(5), Eq.(31), Eq.(32), specially Eq.(33) and the entropy given by Eq.(34) as a direct implication of Eq.(33). So, basing on this great advantage brought by SSR-cosmology to address the creation of the universe, Eq.(34) leads to a null entropy at the Planck scale (origin of the big bang and the dark energy or Planckian vacuum) and can also be extended to the most extreme trans-Planckian regime ({\it nothing}) given to the infinite value of the speed of light $c'=c(\mathcal T_P)=\infty$ (see Eq.(5)) related to the trans-Planckian vacuum pre-existing to creation with null curvature (see Fig.6;a), so that the Planck time $T_P$ would be exactly null for such extreme trans-Planckian regime, as $T_P(c'\rightarrow\infty)=0$ (or $lim_{c\rightarrow\infty}\sqrt{G\hbar/c^5}=0$). Thus, by making $T_P=0$ in Eq.(34), we find an infinite entropy, i.e., $\mathcal S=\infty$. Indeed, this result is the extreme pre-existing condition given by the {\it chaos} or {\it nothing}. 

Now, only according to the cosmology of SSR, it becomes clear to perceive that the process of creation (big bang) from
{\it nothing} represents of starting from an infinite entropy ($\mathcal S=\infty$) of such a
trans-Planckian extreme condition ({\it chaos} or {\it nothing} for $T_P=0$ or $c'\rightarrow\infty$) and going
to the total {\it order} with null entropy ($\mathcal S=0$) at the Planck scale $L_P(=cT_P)$, thus leading
to the big bang (see also Fig.6;c,d). Of course, the infinite decrease of entropy from infinite (chaos) to zero (order), i.e., 
$\Delta S=-\infty$ represents a complete violation of the second law of thermodynamics. Therefore, the probability of transforming the infinite chaos with $\mathcal S=\infty$ ($50\%$ of qubits $(+)$ plus $50\%$ of qubits $(-)$) into a perfect order ($\mathcal S=0$) would be absolutely zero based on the equation of the cosmology of SSR, namely Eq.(34). So we are led to conclude that the infinite decreasing entropy jump is insurmountable only by using probabilistic error-and-hit attempt mechanisms, and so 
{\it chance} is put in doubt by the cosmology of SSR, i.e., the impossibility of occurrence of spontaneous creation. 

Thus we have two possibilities for the cosmology of SSR, namely:

a) Or chaos ({\it nothing}) remains forever, i.e., the cosmic hardware always remains shut down, which is not true because the universe was created with non chance, since the probability for inverting the entropy from infinity ({\it chaos}) to zero ({\it order}) is absolutely zero, which is impossible to happen spontaneously as advocated by the cosmology of spontaneous creation.   

b) Or there must be an {\it information} or program, which is inherent in {\it chaos} or the trans-Planckian vacuum 
({\it nothing}), working like the hardware, thus leading to the big bang ($\mathcal S=0$ by making $m=1$ in Eq.(34)) and the dark energy (Planckian vacuum) without any possibility of error, which destroys the theory of spontaneous creation.  

\subsection{What are qubits in the scenario of quantum cosmology?}

We have seen that time quantization, in the context we are developing, offers a solution to the quantum gravity problem with respect to time. If time is quantized, the same occurs with the spacetime tissue.

The universe began with an original pulse, the first {\it tick-tack} of the big bang cosmic-quantum clock, which occurred in $10^{-43}s$ ($\equiv 1$ qubit). Taking the ultra-referential $S_V$ of SSR, one can calculate the {\it proper time} resulting from the spacetime causality for the low energy region, as well as the variable {\it proper time} recognized as a duration, which 
is equivalent to $\Delta\tau\equiv N(T)$qubit, where $N(T)\equiv 1/\sqrt{1-T_{min}/T}$ and $1~qubit\equiv T_P$. When we achieve much lower temperatures in laboratory, $N(T\rightarrow T_{min}$) increases drastically. This means an increasing of information or 
qubits, thus leading to the increasing of observer's capacity in controlling the quantum collapse like in the Yale experiment. 

SSR has defined a very powerful set of spacetime equations that can be used to solve many physics problems in general. The theory has been very successful in all attempts given in the references quoted in this paper. Applying it in an attempt to understand Yale's experiment served to establish very important points of convergence between Physics and {\it information}. Above all, it was important for understanding new and diverse issues related to time.

Moreover, so we come to realize that the Yale experiment suggests the presence of an {\it information} or program also in charge of the creation starting from a very cold vacuum (not the primordial vacuum with infinite entropy), which just heats to reaching a very high temperature $T_P$, by mantaining a very low entropy. In this sense, an experimental fact as is in Yale adds to our arguments against the spontaneous creation and provides a clue in favor of the suspicion of {\it information} in the big bang control. The big bang would be as if it were a first controlled quantum transition, according to the cosmology of the Symmetrical 
Relativity. Such a very cold vacuum with $T_{min}$ and $S=0$ is between the primordial vacuum ($S=\infty$) and the big bang ($S=0$). Therefore, the non-spontaneous transition of the primordial vacuum (chaos with $T=0$K) to the big bang (Order
with $T=T_P$) passes necessarily by the intermediary cold vacuum (Order with $T_{min}\sim 10^{-12}$K). The future of the universe will recover such a very cold vacuum that will heat again to generate other baby universes by means of other big bangs after its big rip and so on. 

We believe that improving the technique presented in Yale will lead to ever more absolute control of time of collapse and the determinism. So, with such technology, one could build the ideal quantum computer, capable of solving all problems by merging technology and information.  

\subsection{Cosmology of the Symmetrical Relativity and Penrose's Weyl curvature hypothesis}

Penrose's Weyl curvature hypothesis was proposed by him\cite{Penrose1} four decades ago. Since that time, it has provided the motivation for a series of works in General Relativity (GR), so that one has a good knowledge of cosmological models satisfying such hypothesis. More recently, he has described his idea of a conformally-cyclic cosmology in numerous lectures with the fullest account of this circle of ideas so far in his recent work\cite{Penrose2}. 

The starting point for the Weyl curvature hypothesis (WCH) is the observation that the big bang singularity was very special. Penrose argues that the universe must have been initially in a very low entropy state in order for there now to be a second law of thermodynamics. The question is to know why the universe was in a so 
low entropy in the big bang, as its so high temperature $T_P$ would lead to a very high entropy. Penrose 
has searched for the solution of this puzzle by arguing that it would be the foundations of a true quantum 
gravity governed by the asymmetry on time. 

The cosmology of the symmetrical relativity (CSR) contrary to the spontaneous creation seems to provide an explanation for the puzzle of the big bang singularity with a null entropy, since CSR was able to show the decreasing of entropy from infinite in the trans-Planckian regime of {\it nothing} or primordial chaos to the zero entropy in the Planckian regime of the big bang, so that the second law of thermodynamics is satisfied after the big bang according to WCH, where there should be a kind of gravitational entropy.

There is no generally accepted definition of gravitational entropy, but Penrose argues that a low gravitational entropy must mean a small Weyl tensor. But why the Weyl tensor? An answer for this question is that, according to the Belinskii-Khalatnikov-Lifshitz (BKL) singularity picture, singularities of Einstein's vacuum equations, which are necessarily singularities of the Weyl tensor, can be extraordinarily complicated,  probably space-time chaotic. To avoid this chaos, the big bang, while being a cosmological or curvature singularity, needs to have a non-singular Weyl tensor. 

Penrose proposed that the Weyl tensor vanishes at the big bang singularity. A weaker hypothesis which he allowed is simply that the Weyl tensor is finite there. In any way, one has the problem of identifying finiteness of some components of the Riemann tensor, namely those corresponding to the Weyl tensor, simultaneously with other components, belonging to the Ricci tensor, being singular.

One way to deal with this problem is to make a stronger hypothesis, based on the conformal rescaling properties of the curvature. If two metrics $\mathcal G_{ab}$ and $g_{ab}$ are are conformally related, we
write 

\begin{equation}
\mathcal G_{ab}=\Omega^2 g_{ab}, 
\end{equation}
where $\Omega^2$ is a conformal factor of transformation which must obey certain well-known properties. 

Here it is important to call attention to the fact that it was already shown that the metric of SSR 
($\mathcal G_{\mu\nu}$) is a conformal metric\cite{Rodrigo}, so that it was shown that 

\begin{equation}
\mathcal G_{\mu\nu}=\Omega^2 g_{\mu\nu}=\Theta(v)\eta_{\mu\nu}, 
\end{equation}
where we get $\Omega^2=\Theta(v)=1/(1-V^2/v^2)\equiv 1/(1-\Lambda r^2/6c^2)^2$\cite{Rodrigo}, where 
$\Lambda$ represents a cosmological parameter, such that $0<\Lambda<\infty$, and $g_{\mu\nu}=\eta_{\mu\nu}$, which is the Minkowski metric.

Therefore we are led to believe that the cosmology of the symmetrical relativity (SSR) is a promising 
cosmological model which satisfies the Weyl curvature hypothesis, based on the stronger assumption of a cosmological singularity, which is a conformal gauge singularity\cite{Luebble}, since the conformal metric is  regular and the singularity is due to the choice of metric in the conformal class. Singularities like this have also been called isotropic\cite{Goode} and conformally compactifiable\cite{Anguige}. 

It is already known that the Weyl tensor has the same algebraic symmetries as the Riemann tensor by 
obeying the same Bianchi identities. It is generally defined as follows: 

\begin{equation}
W_{\alpha\beta\gamma\delta}=R_{\alpha\beta\gamma\delta}-\frac{1}{(n-2)}(R_{\alpha\gamma}g_{\beta\delta}+R_{\beta\delta}g_{\alpha\gamma}-R_{\alpha\delta}g_{\beta\gamma}-R_{\beta\gamma}g_{\alpha\delta})+
 \frac{1}{(n-1)(n-2)}R(g_{\alpha\gamma}g_{\beta\delta}-g_{\beta\gamma}g_{\alpha\delta}), 
\end{equation}
where $W_{\alpha\beta\gamma\delta}$ is the Weyl tensor, $R_{\alpha\beta\gamma\delta}$ is the Riemann tensor,
$R$ is the scalar curvature and $n$ is the dimensionality of the space.

The main properties of the Weyl tensor that are important for our present analysis are as follows: 

a) It is a traceless tensor, i.e., $g^{\alpha\gamma}W_{\alpha\beta\gamma\delta}=0$.

b) It is conformally invariant, which means that the conformal tensor
$\mathcal W^{\alpha}_{\beta\gamma\delta}=W^{\alpha}_{\beta\gamma\delta}$, only if the following  
conformality condition is obeyed: $\bar g_{\alpha\beta}=\Omega^2g_{\alpha\beta}$. Beyond such essential
condition, we still get $W=0$ in $n\geq 4$. 

Thus, according to those conditions already established above, we conclude that $W=0$ in spacetime of SSR with a given cosmological constant $\Lambda$, since it was already shown before\cite{Rodrigo} that SSR-metric is a conformal metric, i.e., we find $\bar g_{\alpha\beta}\equiv\mathcal G_{\mu\nu}=\Omega^2\eta_{\mu\nu}=\Theta\eta_{\mu\nu}=[1/(1-\Lambda r^2/6c^2)^2]\eta_{\mu\nu}$, where $\Lambda$ is
a positive variable parameter that leads to the current cosmological constant for $r=R_H$ 
(Hubble radius)\cite{N2016}. 

In the big bang, we have found a very high cosmological parameter $\Lambda_P=6c^2/L_P^2$\cite{N2016}, which means that the primordial anti-gravity due to the early dark energy was too strong, thus leading to the 
inflation. As such singularity was governed by a too high cosmological parameter ($\Lambda_P>>0$), the 
temperature was close to the Planck temperature ($\sim 10^{32}K$), so that it was already shown that the scalar curvature $R$ was practically infinite\cite{N2016}\cite{Rodrigo}; however, as SSR-metric is conformally flat even for a very high cosmological parameter $\Lambda_P$, the singularity of the big bang 
had a practically zero Weyl curvature, and so a null gravitational entropy in spite of its too high 
temperature ($T_P$). Therefore, the Planckian regime (big bang regime) of cosmology of the symmetrical relativity (SSR) obeys Penrose's Weyl curvature hypothesis, as the scalar curvature $R$ was too high, but
Weyl curvature $W$ and the entropy $S$ were too small in the big bang, respecting the second law of thermodynamics after big bang. Such issue could be explored deeper elsewhere.  

\section{Conclusion and prospects} 

The cosmology of the Symmetrical Relativity (CSR) represents the proposal of the existence of an implicit determinism as is the proposal of Bohm's theory with respect to the well-known {\it implicit order}. Such a perspective was not part of the initial investigations. We were inclined to advocate only the spontaneous creation, which calls for the easy solution of presenting {\it chance} as the first cause of the big bang, which is now put in doubt by the cosmology of SSR (CSR). 

We realized that the Yale experiment can be adjusted by SSR parameters and have their effects maximized from the theory's perspective, to delve deep into all the issues addressed. In particular, those related to the development of quantum computation, which may reach the highest degree of progress and evolution.

The artificial atom, prepared for use in this first experiment, has a distance of its own and only representative of the actual distance between energy levels of a true atom. The temperature at which the experiment was performed is obviously well above the minimum temperature $T_{min}\sim 10^{-12}$K. And the time involved, which was a few microseconds, is also very far from the
minimum time (Planck time). 

Overcoming the quantization of time due to the Planck time is equivalent to reaching the continuum of infinity. These words would have sounded nonsense before. However, modernity has been exceeding its own expectations of progress because of the speed with which we are witnessing the current changes as the Yale experiment. 

SSR-theory created by Nassif\cite{N2008} is a pioneer of the deformed relativities in the sense that one has never thought of extending relativity theories to the very low speeds and temperatures region to investigate the peculiar causality of minimal energy conditions. So, much more has to be tested in this field of lower energies. 

The article ``Variation of the speed ​​of light and a minimum speed ​​in the scenario of an inflationary universe with accelerated expansion''\cite{TVSL} reveals the possibility of obtaining energy in the ultra-high-vacuum regime. The possibility had not even been raised before. The prospects are really encouraging.

If there is a cosmic program in the context of creation and evolution of the universe - is it an ideal form of AI? Now, AI which one begins to know in modernity, has to be a human achievement.

Deep investigations of nature, of the order found at all scales, of its laws and domains, in view of what we contemplate and are able to understand in the light of geometry, show that we cannot accept {\it chance} of
CEN\cite{CEN} as the origin of the big bang. \\ 

{\noindent\bf  Acknowledgements}

The determinists team seems almost forgotten, despite the great names that compose it as Albert Einstein, David Bohm, Schroedinger, de-Broglie, Roger Penrose {\it et al} to whom we are grateful for the inspiration 
in this work about quantum cosmology, where the Penrose's theorem about the null Weyl tensor in the early
singularity was deeper investigated.

\end{document}